\documentclass[12pt]{article}  
\usepackage[a4paper,textwidth=490.0pt,textheight=703.1pt]{geometry}
\usepackage{dutchcal}
\usepackage{hyperref}
\usepackage{slashed}
\usepackage{graphicx}
\usepackage{amsmath}
\usepackage{amssymb}
\usepackage{color}                                                  
\usepackage{float}
\usepackage[rflt]{floatflt}
\usepackage{cite}
\usepackage{mathtools}
\usepackage{dsfont}
\usepackage{subfig}

\newcommand{\HA}{{\rm H}}
\newcommand{\GA}{{\rm G}}

\usepackage{array}

\usepackage[english]{babel}
\usepackage[latin1]{inputenc}
\usepackage[T1]{fontenc}
\usepackage{ae}

\usepackage{url}

\usepackage{amsmath, amsthm, amssymb}

\newtheorem{thm}{Theorem}[section]

\usepackage{array}

\usepackage[english]{babel}
\usepackage[latin1]{inputenc}
\usepackage[T1]{fontenc}
\usepackage{ae}

\newtheorem{definition}[thm]{Definition}

\newcommand{\Li}{{\rm Li}}
\newcommand{\Mvec}{{\rm \bf M}}

\usepackage{rotating}
\newcommand{\shuffle}{\, \raisebox{1.2ex}[0mm][0mm]{\rotatebox{270}{$\exists$}} \,}
\usepackage{graphicx}

\newcounter{mmacnt}
\def\restartmma{\setcounter{mmacnt}{0}}
\restartmma \catcode`|=\active
\def|#1|{\mathrm{#1}}
\catcode`|=12
\newenvironment{mma}{
 \par\smallskip
 \catcode`|=\active
 \parskip=0pt\parindent=0pt 
 \small
 \def\In##1\\{%
   \def\linebreak{\hfill\break\null\qquad}%
   \refstepcounter{mmacnt}
   \hangindent=2.5em\hangafter=0
   \leavevmode
   \llap{\tiny\sffamily In[\arabic{mmacnt}]:=\kern.5em}%
   \mathversion{bold}\footnotesize$\displaystyle##1$\normalsize
   \mathversion{normal}\par
 }%
 \def\Print##1\\{%
   \def\linebreak{\hfill\break}%
   \hangindent=2.5em\hangafter=0
   \leavevmode ##1\par}%
 \def\Out##1\\{%
   \def\linebreak{$\hfill\break\null\hfill$}%
   \kern\abovedisplayskip\par
   \hangindent=2.5em\hangafter=0
   \leavevmode
   \llap{\tiny\sffamily Out[\arabic{mmacnt}]=\kern.5em}
   \footnotesize$\displaystyle##1$\normalsize\hfill\null\par
   \kern\belowdisplayskip
 }%
 \def\Warning##1##2\\{%
   \def\linebreak{\hfill\break}%
   \hangindent=2.5em\hangafter=0
   \leavevmode
   {\scriptsize##1 : ##2}\par}%
}{%
 \par\smallskip
}

\allowdisplaybreaks[4]

\usepackage{color}

\newenvironment{fshaded}{%
\MakeFramed {\FrameRestore}
}%
{\endMakeFramed}

\usepackage{tikz}
\usetikzlibrary{matrix}

\allowdisplaybreaks[4]

\begin{document}
\setlength{\baselineskip}{0.515cm}
\sloppy
\thispagestyle{empty}
\tikzset{
	graviton/.style={decorate,line width=0.25mm, decoration={snake,amplitude=.5mm, segment length=2mm}},
	massive/.style={postaction={decorate},
		line width=0.4mm,
	},
}
 
\makeatletter
\def\simgt{\mathrel{\lower2.5pt\vbox{\lineskip=0pt\baselineskip=0pt
           \hbox{$>$}\hbox{$\sim$}}}}
\def\simlt{\mathrel{\lower2.5pt\vbox{\lineskip=0pt\baselineskip=0pt
           \hbox{$<$}\hbox{$\sim$}}}}
\makeatother
           
\def\draftnote#1{{\textcolor{red}{\it #1}}}

\def\FT#1{{\color{magenta} [FT: #1]}}
\def\DK#1{{\color{blue} [DK: #1]}}
\def\TS#1{{\color{cyan} [TS: #1]}}

\def\fig#1{Fig.~\ref{#1}}

\def\eqn#1{Eq.~\eqref{#1}}
\def\eqns#1.#2{Eqs.~\eqref{#1} and~\eqref{#2}}
\def\spa#1.#2{\left\langle#1\,#2\right\rangle}
\def\spb#1.#2{\left[#1\,#2\right]}
\def\sand#1.#2.#3{%
\left\langle#1{\vphantom1}\right|{#2}\left|#3\right]}%
\def\sandmp#1.#2.#3{%
\left\langle#1{\vphantom1}\right|{#2}\left|#3\right]}%
\def\sandpm#1.#2.#3{%
\left[#1{\vphantom1}\right|{#2}\left|#3\right\rangle}%
\def\sandmm#1.#2.#3{%
\left\langle#1{\vphantom1}\right|{#2}\left|#3\right\rangle}%
\def\sandpp#1.#2.#3{%
\left[#1{\vphantom1}\right|{#2}\left|#3\right]}%

\def\pp{\sigma}
\def\LL{\mathcal{L}}
\def\phis{\phi_{s}}
\def\CC{C_{2}}
\def\PLS{\mathbb{S}}
\def\Es{\mathcal{E}}
\def\gS{\mathsf{S}}
\def\oS{\mathsf{S}}
\def\sS{\mathsf{S}}
\def\order{O}
\def\op{\mathcal{O}}
\def\opK{\mathcal{K}}

\def\hdelta{{\hat\delta}}
\def\opa{{\hat a}}
\def\opT{{\mathbb{T}}}

\def\opH{\mathcal{H}}
\def\opp{\bm p}
\def\opr{\bm r}
\def\opL{\left(\opr \times \opp\right)}
\newcommand{\spinStr}[1]{\Sigma_{#1}}
\def\clS{\textbf{S}}
\def\clK{\textbf{K}}
\def\KxS{X}
\def\clKxS{\textbf{\KxS}}
\def\eftSigma{\sigma}

\def\doe{\partial}
\def\bs{\boldsymbol}
\def\mc{\mathcal}
\def\clp{\bm p}
\def\clpb{\bar{\bm p}}
\def\clq{\bm q}

\newcommand{\sym}[1]{\{#1\}}

\def\nn{\nonumber}
\def\vs{\vskip 0cm }

\newcommand*\pFqskip{8mu}
\catcode`,\active
\newcommand*\pFq{\begingroup
        \catcode`\,\active
        \def ,{\mskip\pFqskip\relax}%
        \dopFq
}
\catcode`\,12
\def\dopFq#1#2#3#4#5{%
        {}_{#1}F_{#2}\biggl[\genfrac..{0pt}{}{#3}{#4};#5\biggr]%
        \endgroup
}

\newcommand{\MeV}{\rm MeV}
\newcommand{\GeV}{\rm GeV}
\newcommand{\be}{\begin{equation}}
\newcommand{\ee}{\end{equation}}
\newcommand{\eq}[2]{\be\begin{aligned}#1 \label{#2}\end{aligned}\ee}

\newcommand{\Fig}[1]{Fig.~\ref{#1}}
\newcommand{\Eq}[1]{Eq.~\eqref{#1}}
\newcommand{\Eqs}[2]{Eqs.~\eqref{#1} and \eqref{#2}}
\newcommand{\Sec}[1]{Sec.~\ref{#1}}
\newcommand{\Secs}[2]{Secs.~\ref{#1} and \ref{#2}}
\newcommand{\App}[1]{App.~\ref{#1}}
\newcommand{\vev}[1]{\langle #1 \rangle}
\newcommand{\bra}[1]{\langle #1 |}
\newcommand{\ket}[1]{| #1 \rangle}

\newcommand{\sslash}[1]{\ensuremath\raisebox{-0.00cm}{{\small\slash}}\hspace{-0.21cm}#1\/}
\newcommand{\dd}[1]{\frac{\partial}{\partial #1}}

\newcommand{\OurOrder}{  {\cal{O}}(G^3)  }

\newcommand{\ECM}{E_{\rm CM}}
\newcommand{\pCM}{\bm{p}_{\rm CM}}

\newcommand{\mbf}[1]{\mathbf{#1}}

\newcommand{\AEFT}{A_{\rm EFT}}

\newcommand{\MEFT}{M_{\rm EFT}}

\newcommand{\II}{{\cal I}}

\newcommand{\E}{{\rm E}}
\newcommand{\K}{{\rm K}}

\newcommand{\gfkt}[3]{\ell_{#1,#2}^{(#3)}}
\newcommand{\pot}{\rm pot}

\renewcommand{\imath}{\mathrm{i}}

\def\topbotatom#1{\hbox{\hbox to 0pt{$#1\bot$\hss}$#1\top$}} \newcommand*{\topbot}{\mathrel{\mathchoice{\topbotatom\displaystyle} {\topbotatom\textstyle} {\topbotatom\scriptstyle} {\topbotatom\scriptscriptstyle}}}

\newcommand{\tabeq}[2]{ \parbox{#1}{  \be\begin{aligned}#2 \end{aligned} \nonumber \ee }}

\begin{flushleft}
DESY 26--078 \hfill     
\end{flushleft}

\vspace*{2cm}
\begin{center}
{\Large \bf \boldmath The $\mu$-extension of iterated integrals and nested sums}

\vspace*{2mm}
{\Large \bf in quantum field theory}

\vspace*{30mm}
\normalsize
{J. Bl\"umlein$^{a,b}$}, A.M.~Gavrilik$^c$, U.Y.~Lunga$^c$, and O.~Mykhailiv$^c$

\vspace*{5mm}
{\it $^a$Deutsches Elektronen-Synchrotron DESY, Platanenallee 6, 15738 Zeuthen, Germany}

\vspace*{2mm}
{\it $^b$Institut f\"ur Theoretische Physik III, IV, TU Dortmund, \\ Otto-Hahn 
Stra\ss{}e 4, 44227 Dortmund, Germany}

\vspace*{2mm}
{\it $^c$
Bogolyubov Institute for Theoretical Physics, 14-B Metrolohichna str., Kyiv, 03143, Ukraine}
\end{center}

\vspace*{3cm}
\begin{abstract}
  \noindent
  The analytic integration of single-scale Feynman integrals emerging in perturbative 
  calculations in quantum field theories can be performed within special classes of functions,
  which appear as consecutive generalizations of the polylogarithm in form of 
  Kummer-Poincar\'e iterative integrals over special alphabets and extensions thereof. 
  These are the polylogarithms, Nielsen integrals, the iterated integrals over linear 
  denominator terms, cyclotomic letters, letters induced by quadratic forms, and square-root 
  valued letters. These integrals are solutions of first-order factorizing differential 
  equations. They are related to specific nested sums via the Mellin transform and their 
  expansions around $x=0$. We construct the $\mu$-extensions of these iterated integrals and 
  the associated nested sums. We present closed form solutions or provide algorithms in the 
  case of more involved cases to derive the respective $\mu$-extensions and study the 
  algebras of the $\mu$-extended function spaces. Except for the case of square-root 
  valued alphabets, the $\mu$-extension maps into the same function space polynomially in 
  $\mu$. This is also the case for the associated nested sums. For square-root valued 
  alphabets or sums containing central binomials, the $\mu$-extension leads to higher 
  transcendental functions. In all other cases the $\mu$-extension preserves the Hopf algebra 
  structure implied by the (quasi)shuffle product, by supplementing $\mu$ to the ground field.
\end{abstract}


\newpage
\section{Introduction}  
\label{sec:1}

\vspace*{1mm}
\noindent
The construction of $q$-extensions of elementary and special functions, leading to the 
$q$-calculus, has a long history in mathematics, as has been discussed in detail in 
Ref.~\cite{ERNST}. It started with L.~Euler's theory of partitions during the 1740ies 
\cite{EULER1}. Different $q$-analogs have been derived for various special functions, see 
Refs.~\cite{HEINE1,HEINE2,BAILEY,SLATER,EXTON,GASRHA,KOORNWINDER,ANDREWS,KACH,NIST} 
for surveys. They are applied to quantum mechanical systems modifying the Heisenberg
algebra by $q$-deformation \cite{Schwenk:1992sq,Wess:1998ht} 
\begin{eqnarray}
\label{eq1}
\hat{x} \hat{p} - q \hat{p} \hat{x} = i,
\end{eqnarray}
with $q$ the parameter of the extension.\footnote{In some extensions the r.h.s. of Eq.~(\ref{eq1})
is modified by a function of a Hermitian operator.} 
Here $\hat{x}$ and $\hat{p}$ denote the position and the momentum operators. 
The extended functions are related to quantum groups \cite{KASSEL,KS}.\footnote{In the case 
of the $q$-extensions the original field theory is obtained in the limit $q \rightarrow 1$.} 
The $q$-extension
in quantum field theory is implied by 
\begin{eqnarray}
a_\lambda(k) a_{\lambda'}^\dagger(k') - q a^\dagger_{\lambda'}(k') a_\lambda(k)
=  g_{\lambda \lambda'} \delta^{(3)}(k-k'),
\end{eqnarray}
see.~Ref.~\cite{Arefeva:1995rpr,WACHTER}, 
where $a_\lambda(k)$ and $a_{\lambda'}^\dagger(k')$ are the creation and annihilation 
operators.
In this context also concepts of space-time quantization due to non-commutative
geometries are discussed, cf.~Ref.~\cite{CONNES,CONNES1}.\footnote{For early attempts 
see, e.g.,
\cite{Landau:1931nvf,Heisenberg:1936xmg,Snyder:1946qz}.
Snyder's approach Ref.~\cite{Snyder:1946qz} is related, 
cf.~Ref.~\cite{Kowalski-Glikman:2002eyl}, 
to the quantum Poincar\'e group, \cite{Chaichian:1992fx,MANIN,KS},
see also Ref.~\cite{Toller:2003yz}.} 
Different kinds of $q$-extensions are used in the 
literature. In Appendix~\ref{sec:A} we give a brief survey on this.

Another extension consists in the $\mu$-deformation. Here the quantity $\mu$ is 
interpreted as a deformation parameter originally introduced by Jannussis \cite{JANU}, 
see also a more general approach in Ref.~\cite{Gavrilik:2010xu}, in the context of deformations 
of the canonical Heisenberg algebra. In particular, within the framework of the $q$-deformed 
harmonic oscillators with the structure of Lie-admissible algebra 
\cite{ALBERT}, given by commutation relations 
\begin{eqnarray} 
a a^\dagger - a^\dagger a = \varphi_\mu(N+1) - \varphi_\mu(N),~~~~\varphi_\mu(N) 
= \frac{N} {1+\mu N},~~\mu \in [0,1], 
\end{eqnarray} 
where $N$ denotes the number operator and $\mu$ the extension parameter, 
in the quantum mechanical case. For $\mu \rightarrow 0$ the non-deformed theory is 
obtained.
An analogous relation holds for quantum field theories.

The higher functions discussed in the present paper emerge in loop corrections in the non-$q$
or $\mu$-deformed quantum field theories, such as Quantum Chromodynamics or Quantum 
Electrodynamics.
One expects that in the case of the $\mu$-extension the $\mu$-deformation of these 
functions are emerging. 

The $\mu$-deformation has since been employed in various physical contexts, including 
deformed oscillator systems with quasi-Fibonacci spectrum and generalized Bose-gas models 
with  deformed total number of particles,  partition functions, equation of state etc., 
which depend on the parameter $\mu$ \cite{Gavrilik:2010aj,GKR}.  
These constructions modify the thermostatistic properties such as $r$-particle momentum 
correlation functions and virial coefficients. Here $\mu$ can change the inter-particle 
interaction~\cite{Gavrilik:2010xu,Gavrilik:2012ws,Rebesh:2013hps,Gavrilik:2014ooa}. It also 
applies as entropic characteristics of composite boson and interacting 
systems, including the case of treating these 
jointly ~\cite{Gavrilik:2013vza,Gavrilik:2011pz,Gavrilik:2012xr,Gavrilik:2014pxa}.

Furthermore, the $\mu$-deformation has been applied to diverse condensed systems, 
including models of  
galactic dark matter halo structures and induced quantum gravity with 
$\mu$-deformed Einstein 
equations and a $\mu$-dependent effective cosmological constant~\cite{Gavrilik:2014pxa,Gavrilik:2018hoc,
Gavrilik:2018cgj,Gavrilik:2019ccd,Mykhailiv:2025usc,Chung:2018viu,Gavrilik:2023eua}.
 
On the algebraic level, they lead to deformed Heisenberg-type algebras with nontrivial 
operator properties, including extensions exhibiting pseudo-Hermiticity 
and multiparameter generalizations, such as e.g.~($p$,$q$,$\mu$)-deformations, 
cf.~Refs.~\cite{Gavrilik:2012yj,Gavrilik:2023eua}.

In the present paper we consider the $\mu$-extension of higher special functions emerging in 
perturbative quantum field theory. Besides being of mathematical interest, structures of this 
kind may emerge in perturbative calculations in $\mu$-deformed quantum field theories. Both 
in $q$- and $\mu$-extended field theories corresponding extensions of the polylogarithm have 
been already considered, cf.~Refs.~\cite{FS93,Faddeev:1993rs,Rebesh:2013hps}. These 
functions 
form the simplest cases of function spaces studied in the present paper.

Structurally, there is a difference of the function spaces, for which we are constructing
the extension in the following and those, which were mathematically considered in
Refs.~\cite{HEINE1,HEINE2,BAILEY,SLATER,EXTON,GASRHA,KOORNWINDER,ANDREWS,KACH,NIST}. In the 
latter case the determining differential or difference equations are usually of low order, 
while for the functions we are considering the orders are growing with the weight of 
the functions.

The most simple functions appearing in the solution of the differential equations in 
quantum field theory are the logarithm $\ln(1-x)$ \cite{NAPIER1,NAPIER2}, the dilogarithm 
$\Li_2(x)$ \cite{LEIBNIZ1,MAXIMOM}, and the polylogarithms 
$\Li_n(x)$~\cite{SPENCE1,Jonquiere1}, see also \cite{LEWIN1,Devoto:1983tc,LEWIN2}. 
Here we consider the parameter $x \in [0,1]$. The next level of extension consists in the 
Nielsen integrals \cite{NIELSEN1,Kolbig:1983qt}, followed by the harmonic polylogarithms 
\cite{Remiddi:1999ew}. These are special sub-spaces of the generalized harmonic polylogarithms,
which were introduced by E.~Kummer, 
see~Refs.~\cite{KUMMER1,KUMMER2,KUMMER3,POINCARE1,LAPPO,CHEN,GONCHAROV,Moch:2001zr,
Ablinger:2013cf}. Another extension of the harmonic polylogarithms are the cyclotomic harmonic 
polylogarithms \cite{Ablinger:2011te} and the iterated integrals inspired by general 
quadratic forms \cite{Ablinger:2021fnc}.

\begin{table}[H]\centering
{\footnotesize
\renewcommand*{\arraystretch}{1.4}
\textcolor{black}{
\begin{tabular}{|c|r|r|c|}
\hline
\hline
\multicolumn{4}{|c|}{Nested sums}  \\
\hline
\multicolumn{1}{|c|}{Type} &
\multicolumn{1}{c|}{Expression} &
\multicolumn{1}{c|}{Alphabet} &
\multicolumn{1}{c|}{Refs.} \\
\hline
Harmonic sums             & $S_{\vec{a}}(N)$  & $\mathfrak{S}_{H}$   & 
\cite{Vermaseren:1998uu,Blumlein:1998if} \\
\hline
Generalized harmonic sums & $S_{\vec{a}}(\vec{b},N)$ &  $\mathfrak{S}_{GH}$ &
\cite{Moch:2001zr,Ablinger:2013cf}\\
\hline
Cyclotomic harmonic sums  & $S_{\vec{a}}(\vec{b},N)$ & $\mathfrak{S}_{C}$ & \cite{Ablinger:2011te}\\
\hline
Finite binomial sums &    $\binom{2n}{n}^{\pm 1}$ occur in   & $\mathfrak{S}_B$ & 
\cite{Ablinger:2014bra}\\
&    the letters of the alphabet&  &  \\
\hline
\multicolumn{4}{|c|}{Iterated integrals}  \\
\hline
Polylogarithms            & $\Li_n(x)$   & $\mathfrak{A}_P$      & 
\cite{LEWIN1,Devoto:1983tc,LEWIN2}\\
\hline
Nielsen integrals         & $S_{n,p}(x)$ & $\mathfrak{A}_P$ &  \cite{NIELSEN1,Kolbig:1983qt}\\
\hline
Harmonic polylogarithms   & $\HA_{\vec{a}}(x)$ & $\mathfrak{A}_H$ & \cite{Remiddi:1999ew} \\
\hline
Generalized harmonic polylogarithms  &    $\GA_{\vec{a}}(x)$   & $\mathfrak{A}_{KP}$ & 
\cite{KUMMER1,Moch:2001zr,Ablinger:2013cf}\\
\hline
Cyclotomic harmonic polylogarithms  &    $\HA_{\overrightarrow{\{a,b\}}}(x)$   & $\mathfrak{A}_{C}$ & 
\cite{Ablinger:2011te}\\
\hline
Iterative integrals implied &  $\HA_{\overrightarrow{((k,b_k,c_k),d_k)}}(x)$   & $\mathfrak{A}_{R}$ & 
\cite{Ablinger:2021fnc}\\
by quadratic forms          &                       &                     & \\  
\hline
Iterative integrals over &    alphabet contains letters with & 
$\mathfrak{A}_{SQ}$ & 
\cite{Ablinger:2014bra} \\
square-root valued letters    & factors of 
the kind  $(a x^2 + b x +c)^{\pm 1/2}$  & &
\\
\hline
\hline
\end{tabular}}}
\textcolor{black}{
\caption[]{\sf 
The different classes of nested sums and iterated integrals defined in Sections~\ref{sec:3} and 
\ref{sec:4}.
\label{TAB1}
}}
\renewcommand*{\arraystretch}{1.0}
\end{table}

In the previous cases the denominator functions are either linear functions or polynomials. 
Among the iterated integrals, which obey first order factorizing differential equations, 
there are also those derived by square-root valued letters \cite{Ablinger:2014bra}. In 
contrast to special functions which obey second order differential equations or can even 
be obtained as special cases of Gau\ss{}' hypergeometric function, the iterative 
integrals occurring in quantum field theoretic calculations obey higher order
differential equations in general. Likewise, the corresponding nested sums obey higher 
order difference equations. In Table~\ref{TAB1} we summarize these function spaces.

The iterated integrals are related to special types of nested sums via a Mellin transform from $x$-space to 
$N$-space by
\begin{eqnarray}
\label{eq:MEL}
\Mvec[f(x)](N) = \int_0^1 dx~x^{N-1} f(x).
\end{eqnarray}
In this way one obtains the harmonic sums \cite{Vermaseren:1998uu,Blumlein:1998if}, the generalized harmonic sums
\cite{Moch:2001zr,Ablinger:2013cf}, the cyclotomic harmonic sums \cite{Ablinger:2011te}, 
the sums resulting from the functions in Ref.~\cite{Ablinger:2021fnc}, and nested finite 
sums 
containing central binomials $\binom{2n}{n}$ \cite{Ablinger:2014bra}.
These sum-structures do also occur in the Taylor expansion of the iterated integrals at $x=0$.

The easiest way to construct the $\mu$-analog to a given special function $F(x)$ consists in 
determining 
the coefficient function $f[n]$ given by the expansion
\begin{eqnarray}
\label{eq:2}
F(x) = \sum_{n=0}^\infty f[n] x^n.
\end{eqnarray}
Here $f[n]$ has to be known in closed form for each value of $n$. In some cases $F(x)$ 
has logarithmic 
singularities to a finite order at $x=0$ with the representation
\begin{eqnarray}
F(x) = \sum_{m=0}^M \ln^{m}(x) \sum_{n=0}^\infty f_m[n] x^n.
\end{eqnarray}
Also cases with an expansion 
\begin{eqnarray}
F(x) = \sum_{m=0}^M \ln^{m}(x) \sum_{n=0}^\infty f_m[n] x^{n+\alpha},~~~\alpha \in 
\mathbb{R}
\end{eqnarray}
are admissible. For $\alpha = -1/2$ they occur in the case of square-root valued 
alphabets, cf.~Ref.~\cite{Ablinger:2014bra}.

In many cases the functions $f_m[n]$ are the solution of difference equations. If a large 
number of special values of these functions for $n = 0,1,2,...$ can be determined,
one may use guessing-algorithms \cite{GUESS,Blumlein:2009tj,SAGE,GSAGE} to obtain the 
corresponding difference equation
\begin{eqnarray}
\label{eq:REC}
\sum_{l=0}^L p_l(n) f_m[n+l] = 0,
\end{eqnarray}
where $p_l(n)$ are polynomials in $n$.
If Eq.~(\ref{eq:REC}) factorizes to first order, it can be solved by using the algorithms 
available in the package {\tt Sigma} \cite{SIG1,SIG2}. For all the special functions 
given by iterated integrals defined in Section~\ref{sec:21} we will consider in this paper, 
the above steps can be carried out. The functions $f_m[n]$ have a closed form sum-product 
representation, for which one can obtain the $\mu$-analog at least in the form of an
infinite series. In various cases these series can be summed. 
For real representations this is the method of choice, if simpler approaches are not 
available.

The same method can also be applied to calculate the Mellin 
inversion of these iterated integrals $F(x)$, which are given by nested sum-product structures.
In this paper we will derive both the $\mu$-analogs to the different classes of iterated 
integrals and nested sums. 

The paper is organized as follows. In Section~\ref{sec:2} we define the iterated integrals 
and nested sums for which $\mu$-extensions will be derived and discuss their 
(quasi)shuffle 
algebras. In Section~\ref{sec:3} we introduce the iterated integrals for which we 
are going to derive the $\mu$-extension. Likewise, we discuss the different finite nested 
sums appearing in this context in Section~\ref{sec:4}. In Section~\ref{sec:5} we 
summarize the basic operations needed to derive the $\mu$-extensions of the functions 
discussed in Sections~\ref{sec:3} and \ref{sec:4}. The $\mu$-extensions of the nested 
sums are calculated in Section~\ref{sec:6} and the ones of the iterated integrals in 
Section~\ref{sec:7} either in complete form or by devising algorithms by which one 
can obtain the extension in individual cases. 
Section~\ref{sec:8} contains the conclusions. In Appendix~\ref{sec:A} 
we summarize a number of variants of $q$-extensions, which have been proposed in the 
literature. In Appendices~\ref{sec:B} and \ref{sec:C} we describe a series of technical 
aspects.
\section{Basic definitions}  
\label{sec:2}

\vspace*{1mm}
\noindent
In the following we will first define a series of properties of iterated integrals and 
nested sums and some basic operations which are used to get 
the $\mu$-extension of these classes of 
functions.
\subsection{\boldmath Iterated integrals and nested sums}  
\label{sec:21}
	
\vspace*{1mm}
\noindent
Given the alphabet $\mathfrak{A}$ 
\begin{eqnarray}
\mathfrak{A} = \left\{f_1(x), \ldots, f_m(x)\right\},
\end{eqnarray}
where $f_i(x)$ are continuous functions in $x \in ]0,1[$, which are not higher 
transcendental functions.
An iterated integral $G_{a_1,..,a_m}(x)$ over $\mathfrak{A}$ is defined by
\begin{eqnarray}
\label{eq:IT1}
G_{a_1,...,a_m}(x) = \int_0^x dy_1 f_{a_1}(y_1) \int_0^{y_1} dy_2 f_{a_2}(y_2) .... \int_0^{y_{m-1}} 
dy_m
f_{a_m}(y_m).
\end{eqnarray}
Furthermore, we demand that all integrals contributing to (\ref{eq:IT1}) are indefinite, with an 
ordered
sequence of variables $\{x,y_1, \ldots, y_m\}$.
Iterated integrals are the solution of a differential operator which factorizes at first order, with 
\begin{eqnarray}
\frac{d}{dx} G_{a_1,....a_m}(x) = f_{a_1}(x) G_{a_2,....a_m}(x). 
\end{eqnarray}

Likewise, we consider the alphabet $\mathfrak{S}$ 
\begin{eqnarray}
\mathfrak{S} = \left\{s_1(k), \ldots, s_m(k)\right\},~~~~k \in \mathbb{N}.
\end{eqnarray}
Here, $s_l(k)$ are sum-product structures. The nested finite sums are defined by 
\begin{eqnarray}
S_{a,b_1,...,b_m}(N) = \sum_{k=1}^N s_a(k) S_{b_1,...,b_m}(k),
\end{eqnarray}
where the letters $a, b_i$ label elements out of $\mathfrak{S}$.
\subsection{\boldmath Shuffle and quasi-shuffle algebras}  
\label{sec:22}
	
\vspace*{1mm}
\noindent
The notion of alphabets $\mathfrak{S}$ and $\mathfrak{A}$ used below is essential for studying the products of 
iterated integrals and 
nested sums, deriving representations within the respective shuffle and quasi-shuffle algebras 
\cite{Blumlein:1998if,Hoffman:99,Moch:2001zr,Blumlein:2003gb}. 

The products of two iterated integrals $G_{a_1,...,a_l}(x)$ and ${G}_{b_1,...,b_m}(x)$ is the 
sum of all iterated integrals of weight $l+m$ over all shuffles  
\begin{eqnarray}
\{\{c_1,...,c_{l+m}\}\} = \{a_1,...,a_l\} \shuffle \{b_1,...,b_m\}. 
\end{eqnarray}
Here $\shuffle$ denotes the shuffle product.
The indices $a_i, b_j$ refer to the alphabet $\mathfrak{A}_F$ of functions to be iterated.
Here the order of letters of the set $\{a_1,...,a_l\}$ and $\{b_1,...,b_m\}$ is preserved, and 
all other combinations are allowed. One obtains
\begin{eqnarray}
G_{a_1,...,a_l}(x) \cdot G_{b_1,...,b_m}(x) = \sum_{C \in \{\{c_1,...,c_{l+m}\}\}} G_C(x).
\end{eqnarray}

The nested sums obey quasi-shuffle products, since there are additional trace terms, which vanish 
in the case of functions because they are of measure zero. We illustrate the quasi-shuffle 
relations for the product of generalized harmonic sums, which are most important in the following.
The corresponding alphabet is given in Eq.~(\ref{eq:AGH}).
The product of the sums 
$S_{a_1,...,a_l}(c_1,...,c_l;N)$ and $S_{b_1,...,b_m}(d_1,...,d_m;N)$ is obtained by 
\begin{eqnarray}
\label{eq:SHgHS}
\lefteqn{
S_{a_1,...,a_l}(c_1,...,c_l;N) \cdot S_{b_1,...,b_m}(d_1,...,d_m;N) =} \nonumber\\ && 
\hspace*{4cm} \sum_{n=1}^N \frac{c_1^n}{n^{a_1}} S_{a_2,...,a_l}(c_2,...,c_l;n) \cdot 
S_{b_1,...,b_m}(d_1,...,d_m;n)
\nonumber\\ && \hspace*{4cm}
+ \sum_{n=1}^N \frac{d_1^n}{n^{b_1}} S_{a_1,...,a_l}(c_1,...,c_l;n) \cdot 
S_{b_2,...,b_m}(d_2,...,d_m;n) 
\nonumber\\ && \hspace*{4cm}
- \sum_{n=1}^N \frac{(c_1 \cdot d_1)^n}{n^{a_1+b_1}} S_{a_2,...,a_l}(c_2,...,c_l;n) \cdot 
S_{b_2,...,b_m}(d_2,...,d_m;n) 
\end{eqnarray}
iteratively, cf.~Refs.~\cite{Moch:2001zr,Ablinger:2013cf}. The application of 
Eq.~(\ref{eq:SHgHS}) leads 
to a linear combination of generalized harmonic sums with integer coefficients, 
cf.~Ref.~\cite{Ablinger:2013cf}, with additional terms beyond the shuffle-contributions. 

Various explicit examples in the case of 
harmonic sums are given in Refs.~\cite{Blumlein:1998if,Blumlein:2003gb}. One example is
\begin{eqnarray}
S_b(N) \cdot S_{a_1,...,a_l}(N) &=& S_{b,a_1,...,a_l}(N) + S_{a_1,b,a_2,...,a_l}(N) + ... + 
S_{a_1,...,a_l,b}(N) \nonumber\\ &&
- S_{b \wedge a_1,a_2,...,a_l}(N) - ... - S_{a_1,a_2,...,b \wedge a_l}(N),
\end{eqnarray}
with 
\begin{eqnarray}
S_{b,a_1,...,a_l}(N) + S_{a_1,b,a_2,...,a_l}(N) + ... + 
S_{a_1,...,a_l,b}(N) =
S_b(N) \shuffle S_{a_1,...,a_l}(N) 
\end{eqnarray}
and
\begin{eqnarray}
a \wedge b = (|a|+|b|) {\rm sign}(a) {\rm sign}(b).
\end{eqnarray}
The numbers of algebraically independent iterative integrals and nested sums spanning the
respective function space can be calculated by Witt-formulae \cite{WITT1,WITT2} or by 
counting the number of Lyndon words \cite{LYNDON1,LYNDON2,RADFORD}, see also 
Ref.~\cite{Blumlein:2003gb}.
\section{\boldmath Classes of iterated integrals}  
\label{sec:3}
	
\vspace*{1mm}
\noindent
We consider the following classes of iterated integrals: the polylogarithms \cite{LEWIN1,
Devoto:1983tc,LEWIN2}, the Nielsen integrals \cite{NIELSEN1,Kolbig:1983qt}, 
harmonic polylogarithms \cite{Remiddi:1999ew}, generalized harmonic polylogarithms, also
called Kummer-Poincar\'e integrals \cite{KUMMER1,KUMMER2,KUMMER3,POINCARE1,LAPPO,CHEN,
GONCHAROV,Moch:2001zr,Ablinger:2013cf}, cyclotomic iterated integrals \cite{Ablinger:2011te}, 
iterated integrals implied by quadratic forms \cite{Ablinger:2021fnc}, and iterated 
integrals containing square-root valued letters \cite{Ablinger:2014bra}.  

\subsection{\boldmath Polylogarithms}  
\label{sec:31}
	
\vspace*{1mm}
\noindent
Polylogarithms belong to the iterated integrals over the alphabet 
\begin{eqnarray} 
\label{eq:ALPHA:P} 
\mathfrak{A}_P = \left\{\frac{1}{x},\frac{1}{1-x}\right\}. 
\end{eqnarray} 
The second letter only appears as innermost letter, on which the first letter is iterated, see also 
Eq.~(\ref{poly1}).

For $x \in [-1,1]$ the series representation
\begin{eqnarray} 
\label{eq:LIN}
\Li_n(x) = \sum_{k = 1}^\infty \frac{x^n}{k^n},~~n \geq 2,
\end{eqnarray} 
holds. Eq.~(\ref{eq:LIN}) is also the expansion around $x=0$, according to 
Eq.~(\ref{eq:2}).
Furthermore, one has
\begin{eqnarray} 
x \frac{d}{dx} \Li_n(x) = \Li_{n-1}(x),
\end{eqnarray} 
with
\begin{eqnarray} 
\Li_1(x) &=& -\ln(1-x),\\
\Li_0(x) &=& \frac{x}{1-x}.
\end{eqnarray} 
For $n < 1$ the polylogarithms are  rational functions. For $n < 2$ the 
polylogarithm diverges for $x=1$.
\subsection{\boldmath Nielsen integrals}  
\label{sec:32}
	
\vspace*{1mm}
\noindent
Nielsen integrals $S_{n,p}(x)$ \cite{NIELSEN1,Kolbig:1983qt} are all iterated integrals over the alphabet 
(\ref{eq:ALPHA:P}) and cover the classical polylogarithms 
\begin{eqnarray}
S_{n-1,1}(x) = \Li_n(x). 
\end{eqnarray}
They are defined by
\begin{eqnarray}
\label{eq:NIEL1}
S_{n,p}(x) = \frac{(-1)^{n+p-1}}{(n-1)! p!} \int_0^1 \ln^{n-1}(z) \ln^p(1-zx)\frac{dz}{z},~~~n,p \geq 
1,
\end{eqnarray}
see also~Eq.~(\ref{poly2}). The integral-representation (\ref{eq:NIEL1}) shows that $x=1$ is a branch
point both for polylogarithms and Nielsen integrals, at which a cut $[1, \infty)$ starts. 
In the case of the Nielsen integrals one first iterates the 2nd letter in $\mathfrak{A}_P$ $p$ times,
and then the first letter only. Allowing the iteration of both letters in arbitrary order leads to a 
sub-class of the harmonic polylogarithms, cf.~Section~\ref{sec:33}.

The hierarchy of Nielsen integrals is implied by 
\begin{eqnarray}
\label{di1}
x \frac{d}{dx} S_{n,p}(x) &=& S_{n-1,p}(x), \\
\label{int1}
S_{n,p}(x) &=& \int_0^x S_{n-1,p}(y) \frac{dy}{y}.
\end{eqnarray}
The differential relation allows also to define $S_{n,p}(x)$ for $n < 1$, unlike Eq.~(\ref{eq:NIEL1}).
Later on also the Nielsen integrals
\begin{eqnarray}
S_{0,p}(x) &=& \frac{(-1)^p}{p!} \ln(1-x)^p, \\
S_{-1,p}(x) &=& \frac{(-1)^{p-1}}{(p-1)!} \frac{x}{1-x} \ln(1-x)^{p-1},
\end{eqnarray}
contribute. They are diverging at $x=1$.

The expansion of Nielsen integrals around $x = 0$ can be obtained in closed form, 
\begin{eqnarray}
\label{eq:SPN1}
S_{n,2}(x) &=& \sum_{k=2}^\infty S_1(k-1)     \frac{x^k}{k^{n+1}},\\
S_{n,3}(x) &=& \frac{1}{2!} \sum_{k=3}^\infty \left[S_1^2(k-1) - S_2(k-1)\right]  
\frac{x^{k}}{k^{n+1}},
\\   
\label{eq:SPN2}
S_{n,4}(x) &=& \frac{1}{3!} \sum_{k=4}^\infty \left[S_1^3(k-1) - 3 S_1(k-1) S_2(k-1) + 2 
S_3(k-1)\right]  
\frac{x^{k}}{k^{n+1}},
\end{eqnarray}
for the lowest indices $p$. 

The expansion coefficients in 
Eqs.~(\ref{eq:SPN1}--\ref{eq:SPN2}) are related
to Stirling numbers of the 1st kind \cite{COMTET,Kolbig:1983qt,ADAMCHIK1}. 
They can be written  
as modified Faa di Bruno-Ramanujan 
determinants \cite{FDB,RAMAN,Blumlein:1998if}
\begin{eqnarray}
\label{eq:stir}
\left[\begin{array}{c} k \\ p \end{array} \right]
&=& \frac{1}{(p-1)!} \left| \begin{array}{cccccccc}
S_1(k-1) & 1     & 0 &      0 & 0  & ... & 0\\
S_2(k-1) & S_1(k-1)& 2 &      0 & 0  & ... & 0\\
S_3(k-1) & S_2(k-1)& S_1(k-1) & 3 & 0  & ... & 0\\
\vdots &       &        &   &    &     & \vdots  \\
S_{p-1}(k-1) & ...      &        &   &    & ... &S_1(k-1)  \\
\end{array} \right|.
\end{eqnarray}

One obtains
\begin{eqnarray}
\label{eq:SPN3}
S_{n,p}(x) &=& \sum_{k=p}^\infty \left[\begin{array}{c} k \\ p \end{array} \right]    
\frac{x^k}{k^{n+1}},
\end{eqnarray} 
providing the representation of Eq.~(\ref{eq:2}). Note that in the hierarchy--equations
(\ref{di1}, \ref{int1}) the harmonic sums structures, cf.~Section~\ref{sec:41},
 are not 
affected, which is
important for the $\mu$-extension in Section~\ref{sec:72}.
\subsection{\boldmath Harmonic polylogarithms}  
\label{sec:33}
	
\vspace*{1mm}
\noindent
The harmonic polylogarithms $\HA_{\vec{a}}(x)$ are iterated integrals (\ref{eq:IT1}) over the alphabet 
\cite{Remiddi:1999ew}.
\begin{eqnarray}
\mathfrak{A}_{\rm H} = 
\left\{f_0(x), f_1(x), f_{-1}(x)\right\} \equiv 
\left\{\frac{1}{x}, \frac{1}{1-x}, \frac{1}{1+x}\right\},
\end{eqnarray} 
with 
\begin{eqnarray}
\label{int3}
\HA_{b,\vec{a}}(x) = \int_0^x dy f_b(y) \HA_{\vec{a}}(y)
\end{eqnarray} 
and
\begin{eqnarray}
\label{di3}
\frac{d}{dx} \HA_{b,\vec{a}}(x) = f_b(x) \HA_{\vec{a}}(x).
\end{eqnarray} 
With longer and longer alphabets the expansions around $x = 0$ lead to more and more involved explicit structures.

The simplest harmonic polylogarithms are
\begin{eqnarray}
\HA_0(x) &=& \ln(x), \\
\label{eq:lo1}
\HA_1(x) &=& -\ln(1-x) = \sum_{k=1}^\infty \frac{x^k}{k},\\
\label{eq:lo2}
\HA_{-1}(x) &=& \ln(1+x)  = - \sum_{k=1}^\infty \frac{(-x)^k}{k}. 
\end{eqnarray} 

The general series-representation of a specific harmonic polylogarithm can be obtained
by the function {\tt GetMoment} of the package {\tt HarmonicSums} 
\cite{Ablinger:2013eba,Ablinger:2014rba,Ablinger:2010kw,Ablinger:2013hcp,ALL2016,
Ablinger:2015gdg,Ablinger:2018cja,Ablinger:2019mkx,ALL2018,
Blumlein:2009ta,Vermaseren:1998uu,Blumlein:1998if,
Remiddi:1999ew,Ablinger:2011te,Ablinger:2013cf,Ablinger:2014bra,Blumlein:2003gb,
Blumlein:2009cf,Ablinger:2013jta}.\footnote{If not noted otherwise all commands 
quoted in the following are those of the package {\tt HarmonicSums}.} The expansion 
coefficients $f[n]$ of all harmonic  
polylogarithms around $x=0$ are polynomials of nested harmonic sums and rational function of $n$,
including factors  of $(-1)^n$. In the calculations performed in the present paper it is necessary
to perform some of the infinite sums first as finite sums to an upper quantifier $N$ by  
package {\tt Sigma} \cite{SIG1,SIG2} and afterwards to perform the limit $N \rightarrow \infty$
using the routines {\tt SigmaReduce} and {\tt SigmaLimit}.\footnote{Commands in {\tt Sigma} and
{\tt HarmonicSums} may require to newly define the sum type, which can be obtained
by {\tt ToHarmonicSumsSum} prior to {\tt TransformToSSums}.} 

We consider an example at weight {\sf w=6},
\begin{eqnarray}
\label{eq:HPL1}
\HA_{-1, 0, 1, 0, 1, -1}(x) &=& 
\sum_{n=1}^\infty \Biggl\{\Biggl[
        -\frac{1}{n^6}
        +\frac{S_5({n})}{n}
        -\frac{S_{-2,-3}({n})}{n}
        -\frac{S_{-4,-1}({n})}{n}
        +\frac{S_{-2,2,-1}({n})}{n}
\Biggr] (-1)^n
\nonumber\\ &&
+\frac{S_{-1}({n})}{n^5}
+\frac{S_{-3}({n})}{n^3}
-\frac{S_{2,-1}({n})}{n^3} \Biggr\} x^n.
\end{eqnarray} 
One obtains this result by 
\begin{eqnarray} 
{\tt GetMoment[H[-1, 0, 1, 0, 1, -1,x],x,n]}.
\end{eqnarray} 
Differentiating Eq.~(\ref{eq:HPL1}) for $x$ 
\begin{verbatim}
HarmonicSumsSum[D[GetMoment[H[-1,0,1,0,1,-1,x],x,n]*x^n,x],{n, 1, Infinity}] 
// TransformToSSums // SinfToH // VariableToArgument
\end{verbatim}

\noindent
one obtains 
\begin{eqnarray} 
\frac{\HA_{0,1,0,1,-1}(x)}{1+x}
\end{eqnarray} 
as expected, with
\begin{eqnarray} 
\frac{\HA_{0,1,0,1,-1}(x)}{1+x} = \sum_{n=1}^\infty (-1)^n \left[
- S_5(n) + S_{-4,-1}(n) + S_{-2,-3}(n) - S_{-2,2,1}(n) \right] x^n. 
\end{eqnarray} 

The classical polylogarithms and the Nielsen integrals form subsets of the harmonic 
polylogarithms with
\begin{eqnarray} 
\label{poly1}
\Li_n(x)    = \HA_{\tiny {\underbrace{0,...,0}_{n-1},1}}(x),
\\
\label{poly2}
S_{n,p}(x) = \HA_{\tiny {\underbrace{0,...,0}_n,\underbrace{ 
1,...,1}_p}}(x),
\end{eqnarray} 
from which one may also obtain the relations 
Eqs.~(\ref{eq:SPN1},\ref{eq:SPN2},\ref{eq:SPN3}). However, both the polylogarithms and 
Nielsen integrals do not form shuffle algebras, since their products are in general not 
contained in these subsets. The harmonic polylogarithms form a shuffle algebra, 
cf.~Section~\ref{sec:22}.
\subsection{\boldmath Generalized harmonic polylogarithms}  
\label{sec:34}

\vspace*{1mm}
\noindent
Generalized harmonic polylogarithms, also called Kummer-Poincar\'e iterated integrals, 
Refs.~\cite{KUMMER1,KUMMER2,KUMMER3,POINCARE1,LAPPO,CHEN,GONCHAROV,Moch:2001zr,
Ablinger:2013cf}, are iterated 
integrals over the alphabet
\begin{eqnarray} 
\mathfrak{A}_{KP} = \left\{\frac{1}{x-c}\right\},~~~~c \in \mathbb{C}.
\end{eqnarray} 
They are given by
\begin{eqnarray} 
\label{eq:GHPL2}
{\rm G}_{b,\vec{a}}(x) = \int_0^x dy f_b(y) {\rm G}_{\vec{a}}(y),~~~f_c(x) 
\in \mathfrak{A}_{KP}.
\end{eqnarray} 
Since ${\rm G}_{b,\vec{a}}(x)$ may have poles in the integration regions, these have to be removed.\footnote{
This is possible using the command {\tt GLRemovePole[f(x),$x_0$]} of the package {\tt HarmonicSums}, 
with $x_0$ the pole-position.}
The harmonic polylogarithms form a subset of the generalized harmonic polylogarithms.
In the following we will use the symbol {\rm H} also for {\rm G}.

The expansion of generalized harmonic polylogarithms around $x=0$ can 
be carried out in the same  way as in Section~\ref{sec:33}. 
We illustrate this in the example of a weight {\sf w=4} function
$\HA_{-1/2, 2, -1, 2/3}(x)$. The command {\tt GetMoment[H[-1/2, 2, -1, 2/3, x], x, n]}
yields
\begin{eqnarray} 
\label{gqw:GHPL1}
\HA_{-1/2, 2, -1, 2/3}(x) &=& \sum_{n=1}^\infty \Biggl\{
-\frac{\big(
        \frac{3}{2}\big)^n}{n^4}
+\frac{(-1)^n S_1\big({{-\frac{3}{2}},n}\big)}{n^3}
+\frac{2^{-n} S_2({{3},n})}{n^2}
+\frac{(-2)^n S_3\big({{-\frac{3}{4}},n}\big)}{n}
\nonumber\\ &&
-\frac{2^{-n} S_{1,1}\big({{-2,-\frac{3}{2}},n}\big)}{n^2}
-\frac{(-2)^n S_{1,2}\big({{-\frac{1}{4},3},n}\big)}{n}
-\frac{(-2)^n S_{2,1}\big({{\frac{1}{2},-\frac{3}{2}},n}\big)}{n}
\nonumber\\ &&
+\frac{(-2)^n 
S_{1,1,1}\big({{-\frac{1}{4},-2,-\frac{3}{2}},n}\big)}{n}
\Biggr\} x^n.
\end{eqnarray} 
We check now the differentiation relation, based on Eq.~(\ref{eq:GHPL2}). By applying the
same computational steps as in the last section one obtains
\begin{eqnarray} 
\frac{d}{dx} \HA_{-1/2, 2, -1, 2/3}(x) &=& 
\sum_{n=1}^\infty
2 (-2)^n \Biggl[-S_3\left({{-\frac{3}{4}},n}\right)
-S_{1,1,1}\left({{-\frac{1}{4},-2,-\frac{3}{2}},n}\right)
+S_{1,2}\left({{-\frac{1}{4},3},n}\right)\nonumber\\ &&
+S_{2,1}\left({{\frac{1}{2},-\frac{3}{2}},n}\right)\Biggr] x^n.
\end{eqnarray} 
The summation yields
\begin{eqnarray} 
\label{eq:GPL3}
\frac{d}{dx} \HA_{-1/2, 2, -1, 2/3}(x) = \frac{2}{1+2 x} \HA_{2/x,-1/x,2/(3x)}(1)  
= \frac{1}{\frac{1}{2} +x} \HA_{2, -1, 2/3}(x).
\end{eqnarray} 
Here one is advised to work term by term. The last step in Eq.~(\ref{eq:GPL3}) is 
performed by the command {\tt GLVariableToArgument} 

\subsection{\boldmath Cyclotomic harmonic polylogarithms}  
\label{sec:35}
	
\vspace*{1mm}
\noindent

\vspace*{1mm}
\noindent
The cyclotomic harmonic polylogarithms \cite{Ablinger:2011te} are iterative integrals over the 
alphabet
\begin{eqnarray} 
\mathfrak{A}_C = \Biggl\{\frac{p_{i,j}(x)}{C_i(x)}\Biggr\},
\end{eqnarray} 
where $C_i(x)$ is the $i$th cyclotomic polynomial
\begin{eqnarray} 
\Biggl\{
1-x, 1+x, 1+x+x^2, 1+x^2, 1+x+x^2+x^3+x^4, 1-x+x^2, ...
\Biggr\}
\end{eqnarray} 
and
\begin{eqnarray} 
p_{i,j} \in \Biggl\{1,1,\{1,x\},\{1,x\},\{1,x,x^2,x^3\},\{1,x\},...
\Biggr\}.
\end{eqnarray} 
The cyclotomic polynomials are the real factors of
\begin{eqnarray} 
(1-x^n),~~~n \in \mathbb{N},~~n \geq 1,
\end{eqnarray} 
counting the respective new polynomials factors.

This class of iterated integrals is already very involved. Individual expansions 
around $x = 0$ can be obtained by computer algebraic codes, however. To obtain the closed
form solution of $f[n]$ one guesses the corresponding difference equation by using the 
algorithm of Refs.~\cite{GUESS,Blumlein:2009tj,SAGE,GSAGE}.
One obtains a first-order factorizing difference equation 
which can be solved by the package {\tt Sigma} \cite{SIG1,SIG2} to obtain the corresponding closed
form representation. This will involve cyclotomic harmonic sums.

Let us illustrate this in the following example. We consider the real-valued cyclotomic
harmonic polylogarithm
\begin{eqnarray} 
\HA_{\{6,1\},\{1,0\},\{4,0\}}(x) 
\equiv \int_0^x dx_1 \frac{x_1}{1-x_1+x_1^2} \int_0^{x_1} dx_2 \frac{1}{-1+x_2} \int_0^{x_2} 
dx_3 \frac{1}{1+x_3^2},
\label{eq:cycl1}
\end{eqnarray} 
which contains letters of cyclotomy 6, cyclotomy 1, and cyclotomy 4.
To obtain a closed-form series expansion $f[n]$ we have to change to a complex-valued 
representation\footnote{Note that not all commands of {\tt HarmonicSums} are compatible 
with complex-valued operations.}
\begin{eqnarray} 
\lefteqn{\HA_{\{6,1\},\{1,0\},\{4,0\}}(x) =} \nonumber\\ && -\frac{1}{2 \big(
        1+\sqrt[3]{-1}\big)} \Biggl[i \left[
 \text{G}\left(
        \frac{1}{(-1)^{1/3}-\tau },\frac{1}{1-\tau },\frac{1}{i-\tau };x\right)
+ \text{G}\left(
        \frac{1}{(-1)^{1/3}-\tau },\frac{1}{1-\tau },\frac{1}{i+\tau \
};x\right)\right]
\nonumber\\ &&
-(-1)^{5/6} \left[\text{G}\left(
        \frac{1}{(-1)^{2/3}+\tau },\frac{1}{1-\tau },\frac{1}{i-\tau \
};x\right)
+ \text{G}\left(
        \frac{1}{(-1)^{2/3}+\tau },\frac{1}{1-\tau },\frac{1}{i+\tau \
};x\right) \right] \Biggr\}.
\end{eqnarray} 
Here G denotes an iterative integral, giving the list of letters.
The commands {\tt GLToH} and {\tt GetMoment} yield the desired result
\begin{eqnarray}
f[n] &=& 
\frac{1}{2 \big(
        1+\sqrt[3]{-1}\big) n^3} \Biggl\{
        (-i)^{1+n}
        +i^{1+n}
        -(-1)^{5/6} (-i)^n
        +(-1)^{5/6} i^n
\nonumber\\ &&      
  +\big(
                i+\sqrt[6]{-1}
        \big)
\big(2-\sqrt[3]{-1}+2 (-1)^{2/3}\big) n S_1(-i,n)
\nonumber\\ &&       
 -\big(
                \big(
                        i+\sqrt[6]{-1}
                \big)
\big(2-\sqrt[3]{-1}+2 (-1)^{2/3}\big) n S_1({{i},n})\big)
        +n^2 (-1)^{\frac{1}{6} (5+2 n)} \
S_2\big({{-\sqrt[6]{-1}},n}\big)
\nonumber\\ &&  
       -n^2 (-1)^{\frac{1}{6} (5+2 n)} S_2\big({{\sqrt[6]{-1}},n}\big)
        +i n^2 e^{-\frac{1}{3} i n \pi } S_2\big({{-(-1)^{5/6}},n}\big)
        -i n^2 e^{-\frac{1}{3} i n \pi } S_2\big({{(-1)^{5/6}},n}\big)
\nonumber\\ &&     
    -i n^2 e^{-\frac{1}{3} i n \pi } \
S_{1,1}\big({{\sqrt[3]{-1},{-i}},n}\big)
        +i n^2 e^{-\frac{1}{3} i n \pi } \
S_{1,1}\big({{\sqrt[3]{-1},i},n}\big)
        -n^2 (-1)^{\frac{1}{6} (5+2 n)} \
\nonumber\\ && 
S_{1,1}\big({{-(-1)^{2/3},{-i}},n}\big)
        +n^2 (-1)^{\frac{1}{6} (5+2 n)} \
S_{1,1}\big({{-(-1)^{2/3},i},n}\big)
\big)
\Biggr\}, 
\end{eqnarray}
given by generalized sums with complex-valued weights. The first expansion terms read
\begin{eqnarray}
\HA_{\{6,1\},\{1,0\},\{4,0\}}(x) =
\frac{x^4}{8}+\frac{x^5}{6}+\frac{x^6}{12}-\frac{x^7}{35}-\frac{5 x^8}{72}
-\frac{73 x^9}{2835}+\frac{29 x^{10}}{700} + O(x^{11}).
\end{eqnarray}
The derivative of Eq.~(\ref{eq:cycl1}) is given by
\begin{eqnarray}
\lefteqn{\frac{x}{1-x+x^2}
\HA_{\{1,0\},\{4,0\}}(x) = \sum_{n=1}^\infty x^n \frac{1}{2 \big(
        1+\sqrt[3]{-1}\big)} =} \nonumber \\ &&
\times
\Biggl\{
        \big[
                -S_2\big({{-\sqrt[6]{-1}},n}\big)
                -S_{1,1}\big({{-(-1)^{2/3},i},n}\big)
                +S_2\big({{\sqrt[6]{-1}},n}\big)
                +S_{1,1}\big({{-(-1)^{2/3},-i},n}\big)
        \big] (-1)^{n/3}
\nonumber\\ &&
        +\big[
                -S_2\big({{(-1)^{5/6}},n}\big)
                -S_{1,1}\big({{\sqrt[3]{-1},-i},n}\big)
                +S_2\big({{-(-1)^{5/6}},n}\big)
                +S_{1,1}\big({{\sqrt[3]{-1},i},n}\big)
        \big] 
\nonumber\\ &&
\times
(-1)^n \big(
                -\sqrt[3]{-1}\big)^{2 n}
\Biggr\} \sqrt[6]{-1}
\nonumber\\
&=&     \frac{-i x}{2 \big(
        1-x+x^2\big)} \big[
        -S_2({{i x},\infty })
        -S_{1,1}({{x,-i},\infty })
        +S_2({{-i x},\infty })
        +S_{1,1}({{x,i},\infty })
\big].
\end{eqnarray}

\subsection{\boldmath Iterated integrals implied by quadratic forms}  
\label{sec:36}
	
\vspace*{1mm}
\noindent
The alphabet $\mathfrak{A}_R$ for these integrals, cf.~Ref.~\cite{Ablinger:2021fnc}, is given by
\begin{eqnarray}
\mathfrak{A}_R = \mathfrak{A}_{KP} \cup \{
\frac{1}{x^2 + b_i x + c_i},  
\frac{x}{x^2 + b_i x + c_i} \},~~~b_i, c_i \in \mathbb{R}, 4 c_i  >  b_i.
\end{eqnarray}
The notation for the index set is illustrated by 
\begin{eqnarray}
\HA_{((1,b_1,c_1),d_1)}(x) = \int_0^x dy \frac{y^{d_1}}{y^2 + b_1 y + c_1},~~~d_i \in \{0,1\}.
\end{eqnarray}
We consider the example 
$\HA_{
((1,1,2),0),
((0,1,-1),0),
((0,1,1),0)}(x)$, 
for which $f[n]$ shall be derived. 
The original integrals shall be reformulated in terms of general iterative integrals 
first
before one obtains by the methods described before the analytic representation in 
Eq.~(\ref{eq:quad}). By applying the commands
\begin{verbatim}
GetMoment[GL[{1/(VarGL^2 + VarGL - 2), 1/(VarGL + 1), 1/(VarGL - 1)}, x] // 
GLToL // LToH, x, n]//ReduceToBasis
\end{verbatim}
one derives
\begin{eqnarray}
\label{eq:quad}
\HA_{
((1,1,-2),0),
((0,1,-1),0),
((0,1,1),0)}(x) &=& \int_0^x dy_1 \frac{1}{y_1^2+y_1-2} 
\int_0^{y_1} dy_2 \frac{1}{y_2 + 1}
\int_0^{y_2} dy_3 \frac{1}{y_3 - 1} 
\nonumber\\
&=&
\sum_{n=1}^\infty
\frac{1}{3 n} \Biggl[
        \frac{1}{2} S_2({n})
        -\frac{1}{2} S_{-1}({n})^2
\nonumber\\ &&
        -\left(
                -\frac{1}{2}\right)^n S_2({{-2},n})
        +\left(
                -\frac{1}{2}\right)^n S_{1,1}({{2,-1},n}) \Biggr] x^n. \end{eqnarray} 
Finally, we check the differentiation relation for this example. We apply to the 
derivative 
\begin{eqnarray} T_1 = \sum_{n=1}^\infty f[n] n x^{n-1}, \end{eqnarray} 
with $f[n]$ from Eq.~(\ref{eq:quad}), the commands 
\begin{verbatim} {TransformToSSums // SinfToH // VariableToArgument}. 
\end{verbatim} 
This results into  the expected result
\begin{eqnarray}
T_1 = \frac{d}{dx} \HA_{
((1,1,-2),0), 
((0,1,-1),0),  
((0,1,1),0)}(x) = - \frac{\HA_{-1,1}(x)}{x^2 + x -2}. 
\end{eqnarray}
\subsection{\boldmath Iterated integrals over alphabets containing square-root valued letters}  
\label{sec:37}
	
\vspace*{1mm}
\noindent
Nested sums containing central binomials have been studied in 
Ref.~\cite{Ablinger:2014bra}. They are related to iterated integrals with square-root 
valued letters, for which various classes and many examples were given in 
\cite{Ablinger:2014bra}, Eqs.~(3.30--3.61), for the letters of the square-root 
valued letters of the alphabet $\mathfrak{A}_{SQ}$. We first 
derive the function $f[n]$ in these cases.
Let us consider the following iterated integral 
\begin{eqnarray}
\label{eq:T3a}
T_2 = \int_0^x dy_1 \frac{1}{\sqrt{y_1(1-y_1)}}  
      \int_0^{y_1} dy_2 \frac{1}{1+y_2} 
      \int_0^{y_2} dy_3 \frac{1}{\sqrt{1-y_3}}.
\end{eqnarray}
By applying the command 
\begin{center}
{\tt GLToS[$T_2$]}
\end{center}
one obtains
\begin{eqnarray}
\label{eq:T3b}
T_2 &=& \frac{1}{\sqrt{x}} \Biggl\{
\frac{x^3}{5}
+\frac{x^5}{16}
+\sum_{o_1=6}^{\infty } 
\frac{8}{\big(2 o_1 - 1\big)^3}
\Biggl\{
        2^{-2 o_1} \binom{2 o_1}{o_1} \Biggl[
                1
                -\Biggl(
                        \Biggl(
                                2
                                +
                                S_1(o_1)
                                +
                                \sum_{i_1=1}^{o_1} \frac{(-1)^{i_1} 2^{2 i_1}}
{\displaystyle \binom{2 i_1}{i_1} i_1}
\nonumber\\ &&
                                +
                                \sum_{i_1=1}^{o_1} \frac{(-1)^{i_1} 2^{2 i_1} 
                                \sum_{i_2=1}^{i_1} \frac{\displaystyle (-1)^{i_2} 2^{-2 
i_2} \binom{2 i_2}{i_2}}{\displaystyle -1+2 i_2}}{\displaystyle \binom{2 i_1}{i_1} i_1}
                                -2 
                                \sum_{i_1=1}^{o_1} \frac{1}{-1+2 i_1}
                        \Biggr) o_1\Biggr)
       +2 \Biggl(
                        2
                        +
                        S_1(o_1)
\nonumber\\ &&         
                        +
                        \sum_{i_1=1}^{o_1} \frac{(-1)^{i_1} 2^{2 i_1}}{
\displaystyle \binom{2 i_1}{i_1} i_1}
                        +
                        \sum_{i_1=1}^{o_1} \frac{(-1)^{i_1} 2^{2 i_1} 
                        \sum_{i_2=1}^{i_1} \frac{\displaystyle (-1)^{i_2} 2^{-2 i
                        _2} \binom{2 i_2}{i_2}}{\displaystyle -1+2 i_2}}{
\displaystyle
\binom{2 
i_1}{i_1} i_1}
                        -2 
                        \sum_{i_1=1}^{o_1} \frac{1}{-1+2 i_1}
                \Biggr) o_1^2
        \Biggr]
\nonumber\\ &&        
-(-1)^{o_1} \Biggl[
                1+
                \sum_{i_1=1}^{o_1} \frac{\displaystyle (-1)^{i_1} 2^{-2 i_1} \binom{2 
i_1}{i_1}}{-1
                +2 i_1
                }
        \Biggr]
\big(-1+2 o_1\big)
\Biggr\} x^{o_1}
\Biggr\}, 
\end{eqnarray}
from which $f[o_1]$ can be obtained up to a polynomial in $x$ and the overall weight by 
$1/\sqrt{x}$. The reverse command to transform Eq.~(\ref{eq:T3b}) into 
Eq.~(\ref{eq:T3a}) is 
\begin{eqnarray}
{\tt ComputeGeneratingFunction[f[n],x,\{n,6,Infinity\}]}
\end{eqnarray}
for the main contribution and adding the remaining terms.

We check the differential relation, upon which the hierarchy of square-root valued 
iterated integrals is built. With the command {\tt GLToS} applied to 
the derivative Eq.~(\ref{eq:T3b}) we obtain the representation
\begin{eqnarray}
\frac{d}{dx} T_2 = \frac{x^2}{2}
+
\sum_{n=3}^{\infty } x^n \left[
        \frac{2^{1-2 n} \binom{2 n}{n}}{n (-1+2 n)}
        -\frac{1}{n} 2 (-1)^n 
        \sum_{i_1=1}^n \frac{\left( \displaystyle
                -\frac{1}{4}\right)^{i_1} \binom{
                2 i_1}{i_1}}{-1+2 i_1}
\right].
\end{eqnarray}
Using {\tt ComputeGeneratingFunction} results into
\begin{eqnarray}
\label{eq:T3c}
\frac{d}{dx} T_2 = \frac{2}{\sqrt{x(1-x)}}\left[ \int_0^x dy \left(\frac{1}{1+y} 
- \frac{\sqrt{1-y}}{1+y}\right)\right].
\end{eqnarray}
This equals
\begin{eqnarray}
\label{eq:T3d}
\frac{d}{dx} T_2 = \frac{1}{\sqrt{x(1-x)}} \int_0^x dy_1 \frac{1}{1+y_1} \int_0^{y_1} 
dy_2 
\frac{1}{\sqrt{1-y_2}},
\end{eqnarray}
which can be seen differentiating Eqs.~(\ref{eq:T3c}, \ref{eq:T3d}).
\subsection{\boldmath Non-first order factorizing iterated integrals}
\label{sec:38}
	
\vspace*{1mm}
\noindent
In quantum field theoretic calculations also iterated integrals contribute, the alphabets 
of which contain higher transcendental letters. These functions are also called iterating
non-iterative integrals \cite{INII,Behring:2023rlq}. The higher transcendental letters are 
solutions of 
non-factorizing differential equations at least of order two \cite{Broadhurst:1987ei,
Broadhurst:1993mw,Bloch:2013tra,Adams:2015ydq,Remiddi:2016gno,Adams:2017ejb,
Ablinger:2017bjx,Broedel:2019hyg,Blumlein:2019tmi}.
The simplest structures are $_2F_1$-solutions of second order differential equations
which are related to complete elliptic integrals and modular forms~\cite{Ablinger:2017bjx}.
However, as well higher genus solutions contribute. Various of those result from 
Calabi-Yau differential equations, cf., e.g.,~Ref.~\cite{Pogel:2022ken}.
The $\mu$-extension of these integrals requires different techniques than the ones used 
in the present paper. General algorithms to address these in Mellin-space are not yet 
available.
\section{\boldmath Classes of nested sums}  
\label{sec:4}
	
\vspace*{1mm}
\noindent
The iterated integrals discussed in Section~\ref{sec:3} are related to a series of classes of
nested sums by the Mellin transform, Eq.~(\ref{eq:MEL}). We discuss the
following classes of nested sums: the harmonic sums 
\cite{Vermaseren:1998uu,Blumlein:1998if},
the generalized harmonic sums \cite{Moch:2001zr,Ablinger:2013cf}, the cyclotomic harmonic sums
\cite{Ablinger:2011te}, and the binomial harmonic sums \cite{Ablinger:2014bra}.
\subsection{\boldmath Nested harmonic sums}  
\label{sec:41}
	
\vspace*{1mm}
\noindent
The single harmonic sums are given by 
\begin{eqnarray}
S_{b,\vec{a}}(N) = \sum_{m=1}^N \frac{({\rm sign}(b))^m}{m^{|b|}} S_{\vec{a}}(k),~~~~a_i,b \in \mathbb{Z} 
\backslash \{0\},~~~S_\emptyset = 1,
\end{eqnarray}
see Refs.~\cite{Vermaseren:1998uu,Blumlein:1998if}. 
The corresponding alphabet $\mathfrak{S}_H$ is 
\begin{eqnarray}
\mathfrak{S}_{H} = \Biggl\{\frac{({\rm sign}(c))^m}{m^{|c|}}\Biggr\},~~~c \in \mathbb{Z} \backslash 
\{0\},
~~m \in \mathbb{N} \backslash \{0\}.
\end{eqnarray}
\subsection{\boldmath Generalized nested harmonic sums}  
\label{sec:42}
	
\vspace*{1mm}
\noindent
The generalization of nested harmonic sums are implied by the alphabet 
$\mathfrak{S}_{GH}$ 
\begin{eqnarray}
\label{eq:AGH}
\mathfrak{S}_{GH} = \Biggl\{\frac{c^m}{m^{a}}\Biggr\},~~~c \in \mathbb{C} \backslash \{0\},~~m,a \in 
\mathbb{N} \backslash \{0\}.
\end{eqnarray}
These sums are given by
\begin{eqnarray}
S_{b,\vec{a}}(d,\vec{c};N) = \sum_{m=1}^N \frac{d^m}{m^{b}} 
S_{\vec{a}}(\vec{c};k),~~~~
b, a_i \in \mathbb{N} \backslash \{0\},~~~
d, c_i \in \mathbb{C} \backslash \{0\},~~~S_\emptyset = 1,
\end{eqnarray}
cf.~Refs.~\cite{Moch:2001zr,Ablinger:2013cf}. Furthermore, one has
\begin{eqnarray}
S_{0}(\{d\};N) = \sum_{k=1}^N d^k = d \frac{d^N-1}{d-1}.
\end{eqnarray}

\subsection{\boldmath Nested cyclotomic harmonic sums}  
\label{sec:43}
	
\vspace*{1mm}
\noindent
The alphabet for this class, $\mathfrak{S}_C$, is given by, 
\begin{eqnarray}
\mathfrak{S}_C = \left\{\frac{s_1^k}{(a_1 k + b_1)^{c_1}}\right\},
~~~a_i, c_i \in \mathbb{N} \backslash \{0\}, b_i \in 
\mathbb{N}, s_1 =  \pm 1, a_i > b_i.
\end{eqnarray}
cf.~Ref.~\cite{Ablinger:2011te}. The cyclotomic sums are given iteratively by
\begin{eqnarray}
S_{\{a_1,b_1,c_1\},...,\{a_l,b_l,c_l\}}(s_1,...,s_l)(N) 
&=& \sum_{k=1}^N \frac{s_1^k}{(a_1 k + b_1)^{c_1}}
S_{\{a_2,b_2,c_2\},...,\{a_l,b_l,c_l\}}(s_2,...,s_l)(k),~~S_\emptyset = 1.
\nonumber\\
\end{eqnarray}
These are the real representations. In the expansion
of the cyclotomic harmonic polylogarithms around $x=0$ related sums are occurring, which are
generalized harmonic sums with complex-valued numerator weights. These representations 
can be obtained by using the command {\tt CSToRUS}. An example is
\begin{eqnarray}
\lefteqn{\sum_{k=1}^N \frac{S_1(k)}{(-2 + 3k)} =} \nonumber\\ &&
\frac{1}{6 (1+3 N)^2} \Biggl[
        -2 (2+9 N)
        +\big(
                4
                +9 N
                +i \sqrt{3}
                -2 \big(
                        e^{4 \pi i/3}\big)^{3 N}
        \big)
\big(e^{2 \pi i/3}\big)^{3 N}
        +i \big(
                i+\sqrt{3}
        \big)
\big(e^{2 \pi i/3}\big)^{6 N}
\nonumber\\ && 
        +\big(
                4
                +9 N
                -i \sqrt{3}
                -i \big(
                        -i+\sqrt{3}
                \big)
\big(e^{4 \pi i/3}\big)^{3 N}
        \big)
\big(e^{4 \pi i/3}\big)^{3 N}
\Biggr]
-\frac{3 (1+N) S_1({N})}{2 (1+3 N)}
\nonumber\\ && 
+\frac{1}{6 (1+3 N)} \big(
        5
        +9 N
        +2 \big(
                e^{2 \pi i/3}\big)^{3 N}
        +2 \big(
                e^{4 \pi i/3}\big)^{3 N}
\big) S_1({3 N})
+\frac{1}{3} S_{1,1}({3 N})
\nonumber\\ && 
+\left[
        \frac{1
        +\big(
                e^{2 \pi i/3}\big)^{3 N}
        +\big(
                e^{4 \pi i/3}\big)^{3 N}
        }{3 (1+3 N)}
        -\frac{1}{2} \sqrt[3]{-1}
\right] S_1\big({{e^{2 \pi i/3}},3 N}\big)
\nonumber\\ && 
+\left[
        \frac{1
        +\big(
                e^{2 \pi i/3}\big)^{3 N}
        +\big(
                e^{4 \pi i/3}\big)^{3 N}
        }{3 (1+3 N)}
        +\frac{1}{2} (-1)^{2/3}
\right]  S_1\big({{e^{4 \pi i/3}},3 N}\big)
+\frac{1}{3} S_{1,1}\big({{1,e^{2 \pi i/3}},3 N}\big)
\nonumber\\ && 
+\frac{1}{3} S_{1,1}\big({{1,e^{4 \pi i/3}},3 N}
\big)
-\frac{1}{3} \sqrt[3]{-1} S_{1,1}\big({{e^{2 \pi i/3},1},3 N}\big)
-\frac{1}{3} \sqrt[3]{-1} S_{1,1}\big({{e^{2 \pi i/3},e^{2 \pi i/3}},3 N}\big)
\nonumber\\ &&
-\frac{1}{3} \sqrt[3]{-1} S_{1,1}\big({{e^{2 \pi i/3},e^{4 \pi i/3}},3 N}\big)
+\frac{1}{3} (-1)^{2/3} S_{1,1}\big({{e^{4 \pi i/3},1},3 N}\big)
\nonumber\\ &&
+\frac{1}{3} (-1)^{2/3} S_{1,1}\big({{e^{4 \pi i/3},e^{2 \pi i/3}},3 N}\big)
+\frac{1}{3} (-1)^{2/3} S_{1,1}\big({{e^{4 \pi i/3},e^{4 \pi i/3}},3 N}\big).
\end{eqnarray}

\subsection{\boldmath Nested sums containing central binomials}  
\label{sec:44}
	
\vspace*{1mm}
\noindent
The alphabet for this class of sums $\mathfrak{A}_B$, cf.~Ref.~\cite{Ablinger:2014bra}, 
may contain the letters out of the forgoing alphabets which are multiplied by the terms 
\begin{eqnarray}
\frac{1}{4^k} \binom{2k}{k},~~\text{and/or}~~4^k \frac{1}{\displaystyle \binom{2k}{k}}.
\end{eqnarray}
This also applies to sub-sums.

\section{\boldmath The $\mu$-extension: basic relations}
\label{sec:5}

\vspace*{1mm} 
\noindent 
Let us now turn to the $\mu$-extension of the special functions discussed in the previous 
sections.   
We first summarize basic relations occurring in the $\mu$-extension, which form the basic
building blocks to compute the $\mu$-extensions of the iterated integrals and nested sums.

The $\mu$-extension of the number $n$ is given by \cite{REBESH,Mykhailiv:2025usc}
\begin{eqnarray}
\label{eq:mu2}
\{n\}_\mu = \frac{n}{1+ \mu n}.
\end{eqnarray}
This is also sometimes called $\mu$-bracket.
The original functions are obtained in the limit $\mu \rightarrow 0$ for the corresponding 
$\mu$-extensions.
Polynomial terms and summands in the nested sums contain terms of the type
\begin{eqnarray}
\label{eq:nmu}
\frac{1}{\{n\}_\mu^l} = \sum_{k=0}^{l-1} \binom{l}{k} \mu^k \frac{1}{n^{l-k}} + \mu^l. 
\end{eqnarray}
The term $\mu^l$ causes singular contributions in the limit $x \rightarrow 1$ in several cases 
discussed below.

Since sums over the terms Eq.~(\ref{eq:nmu}) will be performed, the term $\mu^l$ will 
introduce {\it integer sums} instead of {\it harmonic sums}, the former ones are 
divergent in the limit $N \rightarrow \infty$. The $\mu$-extension Eq.~(\ref{eq:mu2})
destroys the structural property of harmonic summation by the term $\propto \mu^l$.
This divergence transforms into the one of the $x$-space functions like the
$\mu$-extended polylogarithms, 
Section~\ref{sec:71}, by the contribution of $\Li_0(x)$, and similarly to the other 
functions.

The $\mu$-extension of the factorial $n$ is given by 
\begin{eqnarray}
\{n\}!_\mu = \prod_{k=1}^n \frac{k}{1+ \mu k}.
\end{eqnarray}

The $\mu$-extension of the central binomial $\binom{2n}{n}$ reads 
\begin{eqnarray}
\left. \binom{2n}{n}\right|_\mu = \frac{\{2n\}!_\mu}{(\{n\}!_\mu)^2}.
\end{eqnarray}

The $\mu$-derivative is given by
\begin{eqnarray}
\label{eq:muDERIV}
{\cal D}_x^{(\mu)} = \frac{d}{dx} \left[1 +  \mu x \frac{d}{dx}\right]^{-1}
= \frac{d}{dx}\left[ 1 - \mu \left(x \frac{d}{dx}\right) 
+ \mu^2  \left(x \frac{d}{dx}\right)^2 - \ldots \right] 
\end{eqnarray}
to be understood as expanded series in $\mu$.\footnote{
We note that this $\mu$-deformed derivative was involved in the respective $\mu$-deformed generalization of the 
Lane--Emden equation enabling the successful modeling \cite{Gavrilik:2019ccd} of the rotation curves
of a number of dwarf galaxies. From two distinct (though equivalent) $\mu$-deformed versions of
the Lane--Emden equation, a novel deformation of Heisenberg algebra has been derived leading to
unusual uncertainty relations \cite{Gavrilik:2023eua}.}
One obtains
\begin{eqnarray}
\label{eq:MUder}
{\cal D}_x^{(\mu)} x^n = \{n\}_\mu x^{n-1}.
\end{eqnarray}
The inverse operation is 
\begin{eqnarray}
\label{eq:MUint}
\int_0^x t^{n-1} {\cal d}_t^{(\mu)}=  \frac{x^n}{\{n\}_\mu}.
\end{eqnarray}
\section{\boldmath The $\mu$-extension of nested sums}
\label{sec:6}

\vspace*{1mm} 
\noindent
We first consider the $\mu$-extension of several types of nested sums, which will later
occur in the $\mu$-extension of the different iterated integrals.
These are the harmonic sums, the generalized harmonic sums with complex numerator weights,
and the finite (inverse) binomial sums. 
\subsection{\boldmath $\mu$-extension of harmonic sums}
\label{sec:61}

\vspace*{1mm}
\noindent
The $\mu$-extension of the single harmonic sums are given by
\begin{eqnarray}
\label{eq:Sm}
S_{m}(N;\mu) &=& \sum_{n=1}^N \frac{1}{(\{n\}_\mu)^m} = \sum_{k=0}^m \binom{m}{k} \mu^k S_{m-k}(N), m > 0,
~~m \in \mathbb{N},
\\
S_{-|m|}(N;\mu) &=& \sum_{n=1}^N \frac{(-1)^n}{(\{n\}_\mu)^{|m|}} = \sum_{k=0}^{|m|} \binom{|m|}{k} \mu^k 
S_{-|m|+k}(N), 
m < 0,
~~m \in \mathbb{Z}.
\end{eqnarray}
Here also $S_0(N)$~\footnote{This index is usually not considered for harmonic sums,
but needs to be included here.} 
\begin{eqnarray}
S_0(N) &=& N
\end{eqnarray}
contributes and
\begin{eqnarray}
S_1(N) &\propto& \ln(N) + \gamma_E,~~~\text{for}~~N \rightarrow \infty,
\end{eqnarray}
where $\gamma_E$ is the Euler-Mascheroni constant.
Note that the inverse Mellin transforms of integer powers of $N$ are differential 
operators and
no longer, like in the case of nested sums, regular functions or $+$-distributions, cf., e.g.,~\cite{Blumlein:1998if}.

Since both the harmonic sums $S_1(N)$ and $S_0(N)$ contribute to the $\mu$-extensions of
all single harmonic sums, a definition of $\lim_{N \rightarrow \infty} S_m(N;\mu) = \zeta_m(\mu)$, 
$m \geq 2$ is not possible.

For the nested harmonic sums one obtains  
\begin{eqnarray}
S_{b,\vec{a}}(N;\mu) &=& \sum_{n=1}^N \frac{({\rm sign}(b))^n}{(\{n\}_\mu)^{|b|}} 
S_{\vec{a}}(N;\mu),~~~b,a_i \in \mathbb{Z} \backslash \{0\}. 
\end{eqnarray}
By binomial decomposition also $S_{b,\vec{a}}(N;\mu)$ can be represented
as a polynomial in $\mu$ over nested harmonic sums.

For $\forall a_i > 0$ one obtains
\begin{eqnarray}
S_{\vec{a}}(N;\mu) &=& 
\sum_{\alpha_1=0}^{a_1} \binom{a_1}{\alpha_1} ...
\sum_{\alpha_m=0}^{a_m} \binom{a_m}{\alpha_m} \mu^{\alpha_1 + ... \alpha_m}
S_{a_1 - \alpha_1, ...,a_m-\alpha_m}(N).
\end{eqnarray}
The case of values of $a_i < 0$ will also be handled as for the  generalized harmonic sums.
Furthermore, one has
\begin{eqnarray}
S_{\tiny {\underbrace{0,...,0}_{k}}}(N) = \frac{(N+k-1)!}{(N-1)! k!}, 
\end{eqnarray}
which diverge like $\propto N^k$ for $N \rightarrow \infty$. This behavior translates
into a corresponding singularity for $x \rightarrow 1$ for the associated iterative
integrals.

These representations can be used to express products of $\mu$-extended harmonic sums $S_{\vec{a}}(N;\mu)$
as polynomials in $\mu$ of related nested harmonic sums  $S_{\vec{b}}(N)$. 
The latter ones form a
quasi-shuffle algebra \cite{Blumlein:1998if,Hoffman:99,Moch:2001zr,Blumlein:2003gb} 
and due to the shuffle product a Hopf algebra \cite{REUTENAUER},~see 
Refs.~\cite{HOPF,MILNER,SWEEDLER,Kreimer:1997dp}.
\subsection{\boldmath $\mu$-extension of the Stirling numbers of the first kind}
\label{sec:62}

\vspace*{1mm}
\noindent
The $\mu$-deformed Stirling numbers of the 1st kind emerge in the general representation of the 
Nielsen integrals, Eq.~(\ref{eq:stir}). As we consider them only in this context, the counter $p$
is not $\mu$-extended, but remains an integer. The respective determinant
\begin{eqnarray}
\left[\begin{array}{c} n \\ p \end{array} \right]_{\mu}
&=& \frac{1}{(p-1)!} \left| \begin{array}{cccccccc}
S_1(n-1;\mu) & 1     & 0 &      0 & 0  & ... & 0\\
S_2(n-1;\mu) & S_1(n-1;\mu)& 2 &      0 & 0  & ... & 0\\
S_3(n-1;\mu) & S_2(n-1;\mu)& S_1(n-1;\mu) & 3 & 0  & ... & 0\\
\vdots &       &        &   &    &     & \vdots  \\
S_{p-1}(n-1;\mu) & ...      &        &   &    & ... &S_1(n-1;\mu)  \\
\end{array} \right|
\nonumber\\
\end{eqnarray}
remains the one of a $p \times p$ dimensional matrix. 
The $\mu$-deformed single harmonic sums needed are given in Eq.~(\ref{eq:Sm}).

\subsection{\boldmath $\mu$-extension of generalized harmonic sums}
\label{sec:63}

\vspace*{1mm}
\noindent
The generalized harmonic sums \cite{Ablinger:2013cf} have the following $\mu$-extension 
\begin{eqnarray}
\label{eq:GHS1}
S_{b,\vec{a}}(d,\vec{c},N;\mu) &=& \sum_{n=1}^N \frac{d^n}{(\{n\}_\mu)^{b}}
S_{\vec{a}}(\vec{c},N;\mu),~~~b,a_i \in \mathbb{N} \backslash \{0\}, d, c_i \in
\mathbb{C} \backslash \{0\} \nonumber\\
&=& \sum_{k=1}^b \binom{b}{k} \mu^k \sum_{n=1}^N \frac{d^n}{n^{b-k}} 
S_{\vec{a}}(\vec{c},N;\mu).
\end{eqnarray}
One obtains
\begin{eqnarray}
\label{eq:GHS2}
S_{\vec{a}}(\vec{c},N;\mu) &=&
\sum_{\alpha_1=0}^{a_1} \binom{a_1}{\alpha_1} ...
\sum_{\alpha_m=0}^{a_m} \binom{a_m}{\alpha_m} \mu^{\alpha_1 + ... \alpha_m}
S_{a_1 - \alpha_1, ...,a_m-\alpha_m}(\vec{c},N).
\end{eqnarray}
Again the $\mu$-extended generalized harmonic sums are spanned
by the generalized harmonic sums and their quasi-shuffle algebra implies that they form a Hopf-Algebra.
\subsection{\boldmath $\mu$-extension of finite binomial sums}
\label{sec:64}

\vspace*{1mm}
\noindent
The generalized $\mu$-extended harmonic sums are extended by factors of
\begin{eqnarray}
\label{eq:BINsum}
\left. \binom{2n}{n}\right|_\mu = \frac{\displaystyle \prod_{k=n+1}^{2n}\left(k + 
\frac{1}{\mu}\right)}{\displaystyle \prod_{k=1}^{n}\left(k + \frac{1}{\mu}\right)}
= \frac{\big(
        n + 1
        +\frac{1}{\mu }
\big)_n}{\big(
        1+\frac{1}{\mu }\big)_n},
\end{eqnarray}
which is a (growing) rational function in $\mu$ of numerator and denominator degree $n$.

We consider the most simple sum for which we obtain
\begin{eqnarray}
\sum_{n=1}^N \frac{1}{4^n}
\left. \binom{2n}{n}\right|_\mu 
&=&
\frac{ (1+2 \mu )}{4 (1+\mu )}
\pFq{3}{2}{1~, 
\tfrac{3}{2} + \tfrac{1}{2 \mu}~,
2 + \tfrac{1}{2 \mu}~}{
2 + \tfrac{1}{\mu}~,
2 + \tfrac{1}{\mu}}{1}
-\frac{ 4^{-N} \Gamma \big(
        1+\frac{1}{\mu }\big) \Gamma \big(
        1
        +\frac{1}{\mu }
        +2 N
\big)}{\Gamma \big(
        1
        +\frac{1}{\mu }
        +N
\big)^2} 
\nonumber\\ &&
\times \left[-1 +\pFq{3}{2}{
1,
\tfrac{1}{2} + \tfrac{1}{2 \mu} +N,
1 + \tfrac{1}{2 \mu} +N}{
1 + N + \tfrac{1}{\mu},
1 + N+ \tfrac{1}{\mu}}{1} \right].
\end{eqnarray}
It is evident that the $\mu$-extension results into higher transcendental functions, like 
here generalized hypergeometric functions \cite{BAILEY,SLATER}. The recurrences of these 
functions are not first order factorizable, as also the functions discussed  in 
Section~\ref{sec:38}. This structure will persist for more involved sums. 
In this case the $\mu$-extension moves out of the algebra for the $\mu$-free expressions.

In the present paper, we are focusing on sums, the recurrences of which are first order factorizing.
For small values of $\mu$ one may consider a Taylor expansion for the sums containing central binomials. 
This is discussed in Appendix~\ref{sec:B}. 

\section{The \boldmath $\mu$-extension of iterated integrals}
\label{sec:7}

\vspace*{1mm}
\noindent
The $\mu$-extension of iterated integrals are obtained from their analytic expansion
around $x=0$, Eq.~(\ref{eq:2}). Some of the representations can be given in closed form 
in the general case. For others one can be compute the individual cases algorithmically. 
\subsection{The \boldmath $\mu$-deformation of the polylogarithm}
\label{sec:71}

\vspace*{1mm}
\noindent
One obtains 
\begin{eqnarray}
\Li_k(x;\mu) = \sum_{n=1}^\infty \frac{x^n}{(\{n\}_\mu)^k} = \sum_{l=0}^k \binom{k}{l} 
\mu^l
\Li_{k-l}(x),~~~x \in [-1,1[.
\end{eqnarray}
Examples are
\begin{eqnarray}
\Li_2(x;\mu) &=& \sum_{k=1}^\infty \left(\frac{1+\mu k}{k}\right)^2 x^k = \Li_2(x) + 2 
\mu 
\Li_1(x) + \mu^2 \Li_0(x),\\ 
\Li_1(x;\mu) &=& \sum_{k=1}^\infty \left(\frac{1+\mu k}{k}\right) x^k = \Li_1(x) + \mu 
\Li_0(x),\\
\Li_0(x;\mu) &=& \sum_{k=1}^\infty  x^k = \Li_0(x).
\end{eqnarray}

Because of this, the 
$\mu$-deformed polylogarithm\footnote{Note that a version of the
$\mu$-polylogarithm was used in conjunction with the $\mu$-Bose gas model
\cite{Rebesh:2013hps}. In Ref.~\cite{Gavrilik:2018cgj} the representation $\Li_s(x;\mu) = 
\sum_{k=1}^\infty x^k/k^s/(1+\mu k)$ has been used. The associated $\zeta$-values are 
described in Appendix~\ref{sec:C}.
Also in Ref.~\cite{KOORNWINDER} deviating definitions 
in the case of the $q$-dilogarithm are reported, where not all powers of $n$ are changed 
to $(1-q^n)/(1-q)$.} is a linear function of all classical polylogarithms 
for $k \geq 0$, with a polynomial in $\mu$. In the limit $x \rightarrow 1$, $\Li_k(x;\mu)$ 
diverges like 
\begin{eqnarray}
\label{eq:LImdiv}
\Li_k(x;\mu) \propto \mu^k \frac{x}{1-x} - k \mu^{k-1} \ln(1-x).
\end{eqnarray}
Therefore, one cannot derive the $\mu$-extended $\zeta$-values from this representation in the limit $x 
\rightarrow 1$. Yet another form of a $\mu$-deformed polylogarithm has been introduced in 
Ref.~\cite{Rebesh:2013hps,Gavrilik:2018cgj}.

However, one may also define the $\zeta$-function using the Dirlichet $\eta$-function, 
\begin{eqnarray}
\label{DIRICH1}
\zeta_s = \frac{1}{1-2^{1-s}} \eta_s,~~~~~s \in \mathbb{Z} \backslash \{1\},
\end{eqnarray}
where $\eta_s$ is defined by 
\begin{eqnarray}
\eta_s = \sum_{k=1}^\infty \frac{(-1)^k}{k^s} = \Li_s(-1),
\end{eqnarray}
see Refs.~\cite{DIRICHLET,HARDYR,TITCH1,Blumlein:1998if}. The 
$\mu$-extended $\eta$-function may be obtained as follows. First one calculates the 
$\mu$-extension of $\Li_s(-x)$ for $x > -1$,
\begin{eqnarray}
\Li_s(-x;\mu) = - \sum_{i=0}^s \binom{s}{i} \mu^i \Li_{s-i}(-x),
\end{eqnarray}
with $\Li_0(-x) = x/(1+x)$ and
\begin{eqnarray}
\Li_{s}(-1) = \left(1-\frac{1}{2^{s-1}}\right) \zeta_s,~~~s \geq 2,~~\Li_{1}(-1) = 
\ln(2).
\end{eqnarray}
$\Li_0(-1)$ is now obtained as analytic continuation $\Li_0(-1) = 1/2$, by which
\begin{eqnarray}
\eta_s(\mu) = \Li_{s}(-1;\mu).
\end{eqnarray}

We finally apply the derivative (\ref{eq:muDERIV}) to the $\mu$-extended polylogarithm using 
Eq.~(\ref{eq:MUder}) 
\begin{eqnarray}
\label{eq:LIM1}
x {\cal D}_x^{(\mu)} \Li_m(x;\mu) = \sum_{n=1}^\infty \frac{x^n}{(\{n\}_\mu)^{m-1}}
= \Li_{m-1}(x;\mu).
\end{eqnarray}

\noindent
Eq.~(\ref{eq:LIM1}) defines the set of $\mu$-extended polylogarithms hierarchically.
\subsection{The \boldmath $\mu$-deformation of the Nielsen integrals}
\label{sec:72}

\vspace*{1mm}
\noindent 
By using Eq.~(\ref{eq:SPN3}) and replacing the single harmonic sums and the denominator 
function by their $\mu$-deformed values one obtains the $\mu$-deformed Nielsen integrals
$S_{n,p}(x;\mu)$, 
\begin{eqnarray}
\label{eq:SPN1mu}
S_{n,p}(x;\mu) &=& 
\sum_{k=p}^\infty \left[\begin{array}{c} k \\ p \end{array} 
\right]_\mu \frac{x^k}{\{k\}_\mu^{n+1}}.    
\end{eqnarray} 
An example is
\begin{eqnarray}
S_{n,2}(x;\mu) 
&=& \sum_{k=2}^\infty S_1(k-1;\mu) \frac{x^k}{\{k\}_\mu^{n+1}} \nonumber\\ 
&=& \sum_{k=2}^\infty [S_1(k-1) + \mu (k-1)] \frac{x^k}{\{k\}_\mu^{n+1}}  \nonumber\\
&=& 
\sum_{l=0}^{n+1} 
\binom{n+1}{l} \mu^l 
\sum_{k=2}^\infty [S_1(k-1) + \mu (k-1)] 
\frac{x^k}{k^{n+1-l}}\nonumber\\
&=&
\sum_{l=0}^{n+1} 
\binom{n+1}{l} \mu^l [S_{n-l,2}(x) + \mu(\Li_{n-l}(x) - \Li_{n+1-l}(x))].
\end{eqnarray}
By resumming the 
individual monomials one obtains harmonic polylogarithms over the alphabet 
$\mathfrak{A}_P$ in the order requested by the Nielsen integrals and logarithmic 
derivatives thereof.
We consider the example $S_{2,3}(x;\mu)$, which reads
\begin{eqnarray}
S_{2,3}(x;\mu) &=& -\Biggl\{-\HA_{0,0,1,1,1}(x)
+\mu  \big(
        -\HA_{0,1,1}(x)
        +2 \HA_{0,0,1,1}(x)
        -3 \HA_{0,1,1,1}(x)
\big)
+\mu ^2 \Biggl(
        -\frac{1}{2} \HA_1(x)
\nonumber\\ &&
        -\frac{3}{2} \HA_1^2(x)
        -\frac{1}{2} \HA_1^3(x)
        +\frac{3}{2} \HA_{0,1}(x)
        -\HA_{0,0,1}(x)
        +6 \HA_{0,1,1}(x)
\Biggr)
+\mu ^3 \Biggl(
        - \frac{3 x}{2 (1-x)}
\nonumber\\ &&
        - \frac{3 (-3+5 x) \HA_1(x)}{2 (1-x)}
        -\frac{(-6+7 x) \HA_1(x)^2}{2 (1-x)}
        -3 \HA_{0,1}(x)
\Biggr)
+\mu ^4 \Biggl(
        -\frac{x (-6+11 x)}{2 (1-x)^2}
\nonumber\\ &&
        +\frac{\big(
                -3+7 x-5 x^2\big) \HA_1(x)}{(1-x)^2}
\Biggr)
- \mu^5 \frac{x^3}{(1-x)^3}\Biggr\}. 
\end{eqnarray}
In calculating expressions of this type some of the sums have to be computed using
the command {\tt ComputeGeneratingFunction}, which establishes and solves a differential 
equation for the problem at hand. Also in the case of the Nielsen integrals one
observes a divergent behavior in the limit $x \rightarrow 1$. For $S_{2,3}(x;\mu)$
the contributions from $O(\mu^3)$ on are concerned.

The hierarchy--equations for the $\mu$-extended Nielsen integrals  are 
\begin{eqnarray}
x {\cal D}_x^{(\mu)} S_{n,p}(x;\mu) &=& 
S_{n-1,p}(x;\mu), \\
\int_0^x S_{n-1,p}(t;\mu) \frac{{\cal d}_t^{(\mu)}}{t} &=&  
S_{n,p}(x;\mu).
\end{eqnarray}

\subsection{The \boldmath $\mu$-deformation of other iterated integrals}
\label{sec:73}

\vspace*{1mm}
\noindent
It has been shown in Sections~\ref{sec:31}--\ref{sec:37} that the functions $f[n]$ are polynomials 
of 
\begin{eqnarray}
\frac{1}{\{n\}^l},~~~~~~S_{\vec{a}}(\vec{b},n),~~~l, 
a_i \in \mathbb{N} \backslash 
\{0\},~~b_i \in \mathbb{C} \backslash \{0\}.
\end{eqnarray}
Therefore the $\mu$-extension of $f[n]$, $f[n;\mu]$, can polynomially
be represented by linear combinations of the terms
\begin{eqnarray}
\frac{1}{\{n\}_{\mu}^l} 
= \left(\frac{1}{n} + 
\mu\right)^l,~~~~S_{\vec{a}}(\vec{b},n;\mu),~~~l, 
a_i \in \mathbb{N} \backslash 
\{0\},~~b_i \in \mathbb{C} \backslash \{0\}.
\end{eqnarray}
The functions $f[n;\mu]$ are therefore polynomials in $\mu$, $1/n^m$, and generalized 
$\mu$-extended harmonic sums. Each polynomial coefficient
in $\mu$ of $f[n;\mu] x^n$, $f[n;\mu,k] x^n$, can then be summed over $n \in [1,\infty)$.
The result is a polynomial in $\mu$ and 
generalized harmonic polylogarithms for the classes of harmonic polylogarithms, generalized
harmonic polylogarithms, cyclotomic harmonic polylogarithms and iterative integrals implied 
by quadratic forms.

Let us sum, as an example,
\begin{eqnarray}
T_3 = \sum_{n=1}^{\infty} \frac{S_{-2,1}(n)}{n^2} x^n
\end{eqnarray}
as one of the potential terms.
The command
\begin{verbatim}
(HarmonicSumsSum[S[-2, 1, n]/n^2*x^n, {n, 1, Infinity}] // TransformToSSums 
// SinfToH // VariableToArgument) /. H[a__,-x] :> TransformH[H[a,-x],x] 
// ReduceToHBasis
\end{verbatim}
yields
\begin{eqnarray}
T_3 &=& -\HA_{0,1}(x) \cdot \HA_{0,0,-1}(x)
+ \HA_{0,1}(x) \cdot \HA_{0,-1,-1}(x)
-\HA_{0,0,0,0,-1}(x)
+3 \HA_{0,0,0,1,-1}(x)
\nonumber\\ && 
+3 \HA_{0,0,0,-1,1}(x)
+2 \HA_{0,0,1,0,-1}(x)
-2 \HA_{0,0,1,-1,-1}(x)
-2 \HA_{0,0,-1,1,-1}(x)
-2 \HA_{0,0,-1,-1,1}(x)
\nonumber\\ &&
-\HA_{0,-1,0,1,-1}(x)
-\HA_{0,-1,0,-1,1}(x)
-\HA_{0,-1,-1,0,1}(x)
+\HA_{0,0,0,-1,-1}(x)
+\HA_{0,0,-1,0,1}(x).
\nonumber\\
\end{eqnarray}

\noindent
In the following we consider a number of examples for the different classes of iterated 
integrals.
\subsection{The \boldmath $\mu$-deformation of harmonic polylogarithms}
\label{sec:74}

\vspace*{1mm}
\noindent
We consider the example $\HA_{-1,0,1}(x;\mu)$. The expansion of $\HA_{-1,0,1}(x)$ around $x=0$ 
is given by
\begin{eqnarray}
\HA_{-1,0,1}(x) = \sum_{n=1}^\infty   \left[\frac{1}{n^3} - \frac{(-1)^n}{n} S_{-2}(n)\right]  x^n.
\end{eqnarray}
The $\mu$-extension is obtained by
\begin{eqnarray}
\HA_{-1,0,1}(x;\mu) &=& 
\HA_{-1,0,1}(x)
+\mu  \left(
        \frac{x \HA_{0,1}(x)}{1+x}
        +2 \HA_{-1,1}(x)
\right)
+\frac{1}{2} \mu ^2 \left(
        \frac{(1+5 x) \HA_1(x)}{1+x}
        -\HA_{-1}(x)
\right)
\nonumber\\ &&
+ \frac{\mu ^3 x^2}{1-x^2}.
\end{eqnarray}
To see the hierarchy for $\mu$-extended harmonic polylogarithms we also look at
the derivative by using the series representation
\begin{eqnarray}
{\cal D}_x^{(\mu)} \HA_{-1,0,1}(x;\mu) &=& \sum_{n=1}^\infty  
\frac{n}{1+ \mu n} f[n;\mu] x^{n-1}
= \frac{1}{1+x} \left[ \HA_{0,1}(x) + 2 \mu \HA_1(x) + \mu^2 
\frac{x}{1-x}\right].
\nonumber\\
\end{eqnarray}
This compares to the $\mu$-extension of $\HA_{0,1}(x)/(1+x)$ which is given by
\begin{eqnarray}
\sum_{n=1}^\infty (-1)^n S_{-2}(n;\mu) x^n = \frac{1}{1+x} \left[ \HA_{0,1}(x) + 2 \mu 
\HA_1(x) + \mu^2 \frac{x}{1-x}\right].
\end{eqnarray}
\subsection{The \boldmath $\mu$-deformation of generalized harmonic polylogarithms}
\label{sec:75}

\vspace*{1mm}
\noindent
We consider the $\mu$-extension of the generalized harmonic polylogarithm 
$\HA[-1/2,2,-1,x]$ as a typical example and solve this algorithmically. The associated
function $f[n]$ reads
\begin{eqnarray}
\label{eq:GHS1}
f[n] = \frac{(-1)^n}{n^3}
-\frac{2^{-n}}{n^2} S_1({{-2},n})
-\frac{(-2)^n}{n} S_2\left({{\frac{1}{2}},n}\right)
+\frac{(-2)^n}{n} S_{1,1}\left({{-\frac{1}{4},-2},n}\right).
\end{eqnarray}
It is obtained by applying the command {\tt GetMoment}.
Note that Eq. (\ref{eq:GHS1}) contains a global factor of $1/n$.

Given the structure of the alphabet, we obtain the $\mu$-extension by using {\tt 
ComputeGeneratingFunction}.
\begin{eqnarray}
\label{GHS3}
\left. \HA_{-\frac{1}{2},2,-1}(x)\right|_\mu &=&
\HA_{-\frac{1}{2},2,-1}(x)
+\frac{\mu }{15 (1+2 x)} \Biggl[
        4 (1+2 x) \left(
                \frac{5}{2} \HA_{-\frac{1}{2},2}(x)
                -2 \HA_{-\frac{1}{2},-1}(x)
        \right)
\nonumber\\ &&
        +6 (2+9 x) \HA_{2,-1}(x)
\Biggr]
+\frac{\mu ^2}{15 (-2+x) (1+2 x)} \Biggl[
        3 (-2+x) (1+2 x) \HA_{-\frac{1}{2}}(x)
\nonumber\\ &&  
        +4 (-2+x) (2+9 x) \HA_2(x)
      -10 \big(
                -2-5 x+6 x^2\big) \HA_{-1}(x)
\Biggr]
\nonumber\\ &&
-\frac{2 \mu ^3 x^3}{(-2+x) (1+x) (1+2 x)}.
\end{eqnarray}

The expansion coefficient of the derivative has $2 \HA[2,-1,x]/(1+2x)$ is 
\begin{eqnarray}
f[n] = (-2)^{1+n}\left[
S_{1,1}\left({{-\frac{1}{4},-2},n}\right) -S_2\left({{\frac{1}{2}},n}\right) \right],
\end{eqnarray}
resulting into
\begin{eqnarray}
\label{GHS4}
\left.\frac{2 \HA_{2,-1}(x)}{(1+2x)}\right|_\mu
=
\frac{2}{1+2x} \Biggl[
  \HA_{2,-1}(x) 
+ \mu \frac{2}{3}  \left(
        \HA_2(x)
        -\frac{(1-2 x) \HA_{-1}(x)}{2-x}
\right)
+ \frac{\mu ^2 x^2}{(2-x) (1+x)}
\Biggr].
\end{eqnarray}
The $\mu$-derivative of (\ref{GHS3}) agrees with (\ref{GHS4}). 

\subsection{The \boldmath $\mu$-deformation of cyclotomic harmonic polylogarithms}
\label{sec:76}

\vspace*{1mm}
\noindent
Let us consider the cyclotomic harmonic polylogarithm
\begin{eqnarray}
\text{G}\left[\left\{\frac{1}{1-\tau+\tau^2},\frac{1}{1+\tau}\right\},x\right].
\end{eqnarray}
The associated function $f[n]$ is obtained by
\begin{eqnarray}
f[n] = 
-\frac{(-1)^{2/3} \big(
        -(-1)^{2/3}\big)^n}{\big(1+\sqrt[3]{-1}\big) n}
S_1\big({{-\sqrt[3]{-1}},n}\big)
+\frac{(-1)^{2/3 +n/3}}{\big(1+\sqrt[3]{-1}\big) n}
S_1\big({{(-1)^{2/3}},n}\big).
\end{eqnarray}
The $\mu$-extension $f[n;\mu]$ reads
\begin{eqnarray}
\lefteqn{f[n;\mu] = \frac{\mu ^2}{3} \big(
        \big(
                \big(
                        -1+(-1)^{n/3}
                \big)
\big(3+i \sqrt{3}\big) 2^{2 n}
                -2 i \sqrt{3} \big(
                        i+\sqrt{3}\big)^{2 n}
        \big) i^n} 
\nonumber\\ && 
        +\big(
                \big(
                        3-i \sqrt{3}
                \big)
\big(-1+\big(
                                -(-1)^{2/3}\big)^n\big) 2^n
                +2 i \sqrt{3} \big(
                        1-i \sqrt{3}\big)^n
        \big) (-1)^n \big(
                i+\sqrt{3}\big)^n
\big) 
\nonumber\\ && \times
2^{-1-3 n} \big(
        i+\sqrt{3}\big)^n
\nonumber\\ &&
+\frac{\mu }{3 n} \big(
        \big(
                \big(
                        -3
                        +\big(
                                3+i \sqrt{3}\big) (-1)^{n/3}
                        -i \sqrt{3}
                        +2 i n \sqrt{3} S_1\big({{\sqrt[3]{-1}},n}\big)
                \big) 2^{2 n}
\nonumber\\ && 
 -4 i \sqrt{3} \big(
                        i+\sqrt{3}\big)^{2 n}
        \big) i^n
        +\big(
                \big(
                        -3
                        +i \sqrt{3}
                        +\big(
                                3-i \sqrt{3}
                        \big)
\big(-(-1)^{2/3}\big)^n
\nonumber\\ &&                        
-2 i n \sqrt{3} S_1\big({{-(-1)^{2/3}},n}\big)
                \big) 2^n
                +4 i \sqrt{3} \big(
                        1-i \sqrt{3}\big)^n
        \big) (-1)^n \big(
                i+\sqrt{3}\big)^n
\big) 2^{-1-3 n} 
\nonumber\\ &&
\times \big(
        i+\sqrt{3}\big)^n
+i \frac{1}{n^2} \big(
        \big(
                -\big(
                        i+\sqrt{3}\big)^{2 n}
                +n 2^{2 n} S_1\big({{\sqrt[3]{-1}},n}\big)
        \big) i^n
        +\big(
                \big(
                        1-i \sqrt{3}\big)^n
\nonumber\\ &&               
 -n 2^n S_1\big({{-(-1)^{2/3}},n}\big)
        \big) (-1)^n \big(
                i+\sqrt{3}\big)^n
\big) \frac{2^{-3 n}
 \big(
        i+\sqrt{3}\big)^n}{\sqrt{3}}.
\end{eqnarray}
By the methods described in the previous sections, the $\mu$-extension in $x$-space is 
obtained as
\begin{eqnarray}
\lefteqn{\left.
\text{G}\left[\left\{\frac{1}{1-\tau+\tau^2},\frac{1}{1+\tau}\right\},x\right]
\right|_\mu =} \nonumber \\ &&
\frac{(-1)^{2/3}}{1+\sqrt[3]{-1}} \Biggl[
        \text{G}\left(
                \frac{1}{(-1)^{1/3}+\tau },\frac{1}{1+\tau };x\right)
 -\text{G}\left(
                \frac{1}{-(-1)^{2/3}+\tau },\frac{1}{1+\tau };x\right)
\Biggr]
\nonumber\\ &&
+ \mu\Biggl[
        -\frac{\big(
                1+x^2\big) \HA_{-1}(x)}{1+x+x^2}
        +\frac{\sqrt[6]{-1} \HA_{-1}\big(
                -(-1)^{2/3} x\big)}{\sqrt{3}}
        -\frac{1}{3} \big(
                -1+(-1)^{2/3}
        \big)  \HA_{-1}\big(
                (-1)^{1/3} x\big)
\Biggr]
\nonumber\\ &&
+ \mu^2 \frac{x^2}{(1+x) \big(
        1+x+x^2\big)}.
\end{eqnarray}
The $\mu$-extension of the derivative $\HA_{-1}(x)/(1-x+x^2)$ is given by
\begin{eqnarray}
\left. \frac{\HA_{-1}(x)}{1-x+x^2}\right|_\mu = \frac{\HA_{-1}(x)}{1-x+x^2}
+ \mu \frac{x}{(1+x)(1-x+x^2)}.
\end{eqnarray} 
It agrees with  ${\cal 
D}_x^{(\mu)}\text{G}\left[\left\{\frac{1}{1-\tau+\tau^2},\frac{1}{1+\tau}\right\},x\right]$.
\subsection{The \boldmath $\mu$-deformation of iterated integrals implied by quadratic forms}
\label{sec:77}

\vspace*{1mm}
\noindent
We perform the $\mu$-extension of the iterative integral (\ref{eq:quad}). It reads
\begin{eqnarray}
\lefteqn{\left. \HA_{
((1,1,2),0),
((0,1,-1),0),
((0,1,1),0)}(x)\right|_\mu = }\nonumber\\ &&  
\frac{1}{3} \HA_1(x) \HA_{-1,1}(x)
-\frac{2}{3} \HA_{-1,1,1}(x)
+\frac{1}{3} \HA_{-2,-1,1}(x)
+\mu  \Biggl[
        -\frac{1}{6} \HA_{-1}(x) \HA_1(x)
        +\frac{1}{6} \HA_1(x)^2
\nonumber\\ &&
        +\frac{\big(
                2-4 x-x^2\big) \HA_{-1,1}(x)}{3 (-1+x) (2+x)}
        +\frac{5}{6} \HA_{-2,1}(x)
        -\frac{1}{6} \HA_{-2,-1}(x)
\Bigg] 
\nonumber\\ && 
+\mu ^2 \Biggl[
        -\frac{x}{6 (-1+x)}
        +\frac{\big(
                14-11 x-68 x^2-7 x^3\big) \HA_1(x)}{36 (-1+x) (1+x) \
(2+x)}
        +\frac{\big(
                -2+3 x+x^2\big) \HA_{-1}(x)}{4 (-1+x) (2+x)}
\nonumber\\ &&   
     -\frac{4}{9} \HA_{-2}(x)
\Biggr]
+ \mu^3 \frac{x^3}{(-1+x)^2 (1+x) (2+x)}.
\end{eqnarray}
Here the term $O(\mu^0)$ is written in an equivalent representation to the one in 
Eq.~(\ref{eq:quad}).

The $\mu$-extension of the derivative of (\ref{eq:quad}) reads
\begin{eqnarray}
\left. -\frac{\HA_{-1, 1}(x)}{x^2 + x - 2}\right|_\mu
&=&
-\frac{\HA_{-1,1}(x)}{x^x+x-2}
- \mu \frac{(1-3 x) \HA_1(x)
        - (1+x) \HA_{-1}(x)
}{2 (x^2+x-2) (1+x) }
\nonumber\\ &&
- \mu^2 \frac{x^2}{(1-x^2) (x^2+x-2)}
\end{eqnarray}
It agrees with ${\cal D}_x^{(\mu)} \left. \HA_{
((1,1,2),0),
((0,1,-1),0),
((0,1,1),0)}(x)\right|_\mu$.
\subsection{\boldmath $\mu$-deformation of iterated integrals with square-root valued
letters}
\label{sec:78}

\vspace*{1mm}
\noindent
The functions $f[n]$ for iterated integrals with square-root valued
letters contains central binomials. Their $\mu$-extension is not polynomial in $\mu$, 
like in the previous cases. The $\mu$-extended central binomials are (growing) 
rational functions in $\mu$. Variable transformations will imply real-valued 
sum-quantifiers. Already in very simple cases
one obtains higher transcendental functions, as the example
\begin{eqnarray}
\sum_{n=1}^\infty \frac{4^{-n} x^n \big(
        1
        +\frac{1}{\mu }
        +n
\big)_n}{\big(
        1+\frac{1}{\mu }\big)_n}
= 
\frac{ (1+2 \mu ) x}{4 (1+\mu )}
\pFq{3}{2}{1~, 
\tfrac{3}{2} + \tfrac{1}{2 \mu}~,
2 + \tfrac{1}{2 \mu}~}{
2 + \tfrac{1}{\mu}~,
2 + \tfrac{1}{\mu}}{x}
\end{eqnarray}
shows. In any case one may write an infinite sum, after the $\mu$-extension of all 
terms occurring in $f[n]$ in the central binomial case. Algorithms, which resum this
representation are not yet known in general.

As an example we consider 
\begin{eqnarray}
T_4 &=& 
\sum_{i=1}^\infty \frac{(4x)^i}{\displaystyle i^2 \binom{2i}{i}} S_1(i)
= \frac{1}{3} x \big(
        6-45 x+88 x^2-42 x^3\big)
+64 
        \text{G}\left(
                \frac{1}{\tau },\sqrt{1-\tau } \sqrt{\tau \
},\sqrt{1-\tau } \sqrt{\tau };x\right)
\nonumber\\ &&
+64 
        \text{G}\left(
                \sqrt{1-\tau } \sqrt{\tau },\frac{1}{1-\tau \
},\sqrt{1-\tau } \sqrt{\tau };x\right)
+24 (1-2 x) \sqrt{(1-x) x} \text{G}\big(
        \sqrt{1-\tau } \sqrt{\tau };x\big)
\nonumber\\ &&
+16 (1-2 x) \sqrt{(1-x) x} \text{G}\left(
        \frac{1}{1-\tau },\sqrt{1-\tau } \sqrt{\tau };x\right)
+32 \text{G}\big(
        \sqrt{1-\tau } \sqrt{\tau },\sqrt{1-\tau } \sqrt{\tau };x\big).
\nonumber\\
\end{eqnarray}
Its $\mu$-extension is given by
\begin{eqnarray}
\left. T_4\right|_\mu = \sum_{i=1}^\infty \frac{(4x)^i (1+\mu i)^2}
{\displaystyle i^2 \left. \binom{2i}{i}\right|_\mu} [S_1(i) + \mu i]
\end{eqnarray}
The derivative of $T_4$ is 
\begin{eqnarray}
\label{eq:DT8}
 \frac{d}{dx} T_4 &=& \frac{1}{x} \Biggl[2 x \big(
        1-3 x+8 x^2-4 x^3\big)
+4 \big(
        3-12 x+8 x^2\big) \frac{\sqrt{x}}{\sqrt{1-x}} \text{G}\big(
        \sqrt{1-\tau } \sqrt{\tau };x\big)
\nonumber\\ &&
+8 \frac{\sqrt{x}}{\sqrt{1-x}} \text{G}\left(
        \frac{1}{1-\tau },\sqrt{1-\tau } \sqrt{\tau };x\right)
+64 \text{G}\big(
        \sqrt{1-\tau } \sqrt{\tau },\sqrt{1-\tau } \sqrt{\tau };x\big)\Biggr].
\end{eqnarray}
By applying (\ref{eq:MUder}) one obtains
\begin{eqnarray}
{\cal D}_x^{(\mu)} \left. T_4\right|_\mu = 
\sum_{i=1}^\infty 4 \frac{(4x)^{i-1} (1+\mu i)}{\displaystyle i
\left. \binom{2i}{i}\right|_\mu} [S_1(i) + \mu i]
\end{eqnarray}
which agrees with the $\mu$-extension of (\ref{eq:DT8}).
For small values of $\mu$, one may perform a Taylor expansion, as lined out in Appendix~\ref{sec:B}. 
\section{Conclusions}
\label{sec:8}

\vspace*{1mm}
\noindent
In quantum field theory several classes of iterative integrals and nested sums play 
a central role in analytic higher order calculations. They
form shuffle or quasi-shuffle algebras or sub-systems of those. By virtue of the 
shuffle product, these algebras are also Hopf algebras \cite{REUTENAUER}. These functions  
obey first-order factorizing differential or difference equations. 

In the present paper
we have derived the $\mu$-extension  for these functions, inspired by Ref.~\cite{JANU}. Except for 
the 
case of nested sums containing central binomial sums and iterated integrals containing
square-root valued letters, the $\mu$-extension maps into the same Hopf algebra 
as the non-deformed functions. In these cases the $\mu$-extension adds lower weight 
functions, up to the weight {\sf w = 0} contributions. The latter terms imply 
the singular behavior at $x=1$ for the iterative integrals and for $N \rightarrow 
\infty$ for the nested sums, beyond logarithmic singularities. The point $x=1$ in the 
non-$\mu$-extended case is already a branch point of the functions. 
In the case of iterated integrals with square-root valued letters the central function 
$f[n]$ contains central binomials. Due to this, their $\mu$-extension implies 
growing rational functions in $\mu$ and $n$. The $x$-representation is then an infinite 
sum over these terms, which forms a higher transcendental 
function. Thus, the 
$\mu$-extension leads out of the space of the non-deformed functions. For small values of 
$\mu$, one may consider the series-expansion in $\mu$, the terms of which belong to the 
non-deformed algebras, while the general solutions do not. 

We have shown that the $\mu$-extensions obey analogous differentiation hierarchies as
the $\mu$-free cases by replacing the differential operator $\frac{d}{dx}$ with the
$\mu$-derivative ${\cal D}_x^{(\mu)}$.

It should be emphasized that the set of the $\mu$-deformed special functions constructed and 
studied in this paper are completely new. We believe that these $\mu$-extensions
should play the same important role as their non-deformed counterparts, in the corresponding
$\mu$-deformed analogs of customary quantum field theory models which are yet to be elaborated.
After being completed, the corresponding results will be presented elsewhere in the future.
This will supplement the existing amount of already developed diverse applications of $\mu$-deformed 
calculus along with its physical realization in the realm of thermo-statistical problems corresponding 
to very distinct scales, from hadronic to galactic ones.

\appendix
\section{\boldmath Different $q$-extensions}
\label{sec:A}

\vspace*{1mm}
\noindent
Related to the $\mu$-extension, there is also a number of different $q$-extensions in 
use. In what follows we summarize some of their basic features for completeness.

\vspace*{2mm}
\noindent
\underline{\sf Asymmetric $q$-extension}
	
\vspace*{1mm}
\noindent
The $q$-extension of a number $a$ is given by
\begin{eqnarray}
\label{eq:q0}
\{a\}_q \equiv \frac{1-q^a}{1-q},~~~q \in \mathbb{C} \backslash \{1\}.
\end{eqnarray}
This approach has been used for the $q$-basic functions in 
Refs.~\cite{HEINE1,HEINE2,BAILEY,SLATER,EXTON,GASRHA,KOORNWINDER,ANDREWS,KACH,NIST}.
In the case $q \in \mathbb{N} \backslash \{0\}$ one has
\begin{eqnarray}
\label{eq:q1}
\{n\}_q \equiv \frac{1-q^n}{1-q} = \sum_{l=0}^{n-1} q^l.
\end{eqnarray}
Eq.~(\ref{eq:q1}) allows to define the $q$-factorial 
\begin{eqnarray}
\{n\}_q! \equiv \prod_{k=1}^n \{k\}_q,~~~\{0\}_q! := 1.
\end{eqnarray}
In various sum-representations integer powers of complex numbers $c$ 
\begin{eqnarray}
c^n,~~~c \in \mathbb{C}
\end{eqnarray}
contribute.
This $n$-dependence is not $q$-extended, like also not for the monomial $x^n$, 
see Eq.~(\ref{eq:2}).

The Jackson $q$-derivative is defined by \cite{JACKSON1,JACKSON2,THOMAE1,THOMAE2}
\begin{eqnarray}
{\cal D}_{q} [f](x) =  \frac{f(x) - f(qx)}{(1-q)x}
\end{eqnarray}
with
\begin{eqnarray}
{\cal D}_{q} [x^\alpha] =  \frac{(1-q^\alpha)}{(1-q)} x^{\alpha-1}.
\end{eqnarray}
for the power function.

\vspace*{2mm}
\noindent
\underline{\sf Symmetric $q$-extension}

\vspace*{1mm}
\noindent
Here the extension of a number $a$ reads, see e.g.~Refs.~\cite{KS,BIEDENHARN,MACFARLANE},
\begin{eqnarray}
\{a\}_q = \frac{q^a - q^{-a}}{q - q^{-1}}.
\end{eqnarray}
The $q$-derivative is given by
\begin{eqnarray}
{\cal D}_q[f](x) = \frac{f(qx) - f(q^{-1}x)}{(q - q^{-1})x}
\end{eqnarray}
and
\begin{eqnarray} 
{\cal D}_q[x^\alpha] = \frac{q^\alpha - q^{-\alpha}}{(q - q^{-1})} x^{\alpha - 1}.
\end{eqnarray} 

\vspace*{2mm}
\noindent
\underline{\sf $q$-extension using two variables}

\vspace*{1mm}
\noindent
This extension has been considered in Refs.~\cite{Chakrabarty,Aric,BURBAN3}. The 
extension of a number $a$ is given by
\begin{eqnarray}
\{a\}_{q,r} = \frac{q^a - r^a}{q - r}.
\end{eqnarray}
The $q,r$-derivative is given by
\begin{eqnarray}
{\cal D}_{q,r}[f(x)] = \frac{f(qx) - f(rx)}{(q - r)x},
\end{eqnarray}
resulting in
\begin{eqnarray}
{\cal D}_{q,r}[x^n] = \{n\}_{q,r} x^{n - 1}.
\end{eqnarray}
Three further generalizations have been discussed in Ref.~\cite{BURBAN}, see also 
Refs.~\cite{CHUNG,BOROZOV,BURBAN1}. 
Note that all the different deformed brackets shown herein may involve, besides a number
$a$, as well as an operator, e.g. the excitation number operator $N$ - in this case the corresponding
`bracket' is meant as formal power series.

\section{\boldmath The $\mu$-expansion for square-root valued letters} 
\label{sec:B}

\vspace*{1mm}
\noindent
The complete representation in the case of contributions due to central binomials 
in $n$-space can be considered for small values of $\mu$ applying a Taylor expansion
in the first few powers,
\begin{eqnarray}
\label{eq:Bin1}
\left. \binom{2n}{n}\right|_\mu &=& 
\binom{2 n}{n}
\Biggl\{
        1
        + 2 \mu  n^2 \big[
                  S_1({n})
                - S_1({2 n})
        \big]
        +\mu ^2 \big[
                n^4 \big(
                        2 S_1({n})^2
                        -4 S_1({n}) S_1({2 n})
                        +2 S_1({2 n})^2
\nonumber\\ &&
                        -2 S_2({2 n})
                        +S_2({n})
                        +\zeta_2
                \big)
                + 2 n^3 \big(
                        - S_1({n})
                        + S_1({2 n})
\big)        
        \big]
\Biggr] + O(\mu^3). 
\end{eqnarray}
There are two new aspects: \\
\hspace*{1cm} {\it i)} also harmonic sums at argument $2n$ contribute.\\
\hspace*{1cm} {\it ii)} positive powers of $n$ appear. 

For {\it i)} one has to synchronize arguments, which will lead to contributions due to 
cyclotomic harmonic sums \cite{Ablinger:2011te}. For the sums $S_1(2n)$ and $S_1(n) 
S_1(2n)$
one obtains
\begin{eqnarray}
S_1(2n) &=&  1
-\frac{1}{1+2 n}
+\frac{1}{2} S_1({n})
+S_{\{2,1,1\}}({n}),
\\
S_1(n) \cdot S_1(2n) &=& -S_1({n})
+\frac{2 n S_1({n})}{1+2 n}
-\frac{1}{2} S_2({n})
+2 S_{\{2,1,1\}}({n})
+S_{1,1}({n})
+S_{\{1,0,1\},\{2,1,1\}}({n})
\nonumber\\ &&
+S_{\{2,1,1\},\{1,0,1\}}({n}),
\\
S_1^2(2n) &=&  
1
+\frac{1}{(1+2 n)^2}
-\frac{2}{1+2 n}
-\frac{S_1({n})}{1+2 n}
+S_{\{2,1,1\}}({n}) S_1({n})
-\frac{1}{4} S_2({n})
+\frac{1}{2} S_{1,1}({n})
\nonumber\\ &&
+2 S_{\{2,1,1\}}({n})
-\frac{2 S_{\{2,1,1\}}({n})}{1+2 n}
-S_{\{2,1,2\}}({n})
+2 S_{\{2,1,1\},\{2,1,1\}}({n})
+S_1({n}),
\end{eqnarray}
by applying the commands {\tt Synchronize} and {\tt LinearExpand}. Here cyclotomic 
sums of the kind
\begin{eqnarray}
S_{\{1,0,1\},\{2,1,1\}}({n}) = \sum_{k=1}^n \frac{1}{k} \sum_{l=1}^k \frac{1}{2 l + 1}
\end{eqnarray}
do appear.

{\it ii)}~Some of the positive powers $n^k$ in (\ref{eq:Bin1}) may contribute under the 
Mellin 
transform
and can be absorbed by
\begin{eqnarray}
\sum_{n=0}^\infty x^n \int_0^1 dy~n~y^{n} f(y) = \int_0^1 dy \frac{x y}{(1-x y)^2} f(y).
\end{eqnarray}
In the general case for factors $n^l$ one obtains
\begin{eqnarray}
\sum_{n=0}^\infty n^l z^n = \left(z \frac{d}{dz}\right)^l \frac{1}{1-z}. 
\end{eqnarray}
By the command {\tt GeneralInvMellin} with the option {\tt ComputeMellinConvolution -> 
true} one can obtain the $n$-space expression as a Mellin transform. Techniques of this 
kind were also used in Refs.~\cite{Blumlein:1998if,Fleischer:1998nb,Davydychev:2003mv,
Weinzierl:2004bn}. The final integral can be performed by the command {\tt GLIntegrate}.

A typical example is given by the term
\begin{eqnarray}
\lefteqn{\sum_{n=1}^\infty \binom{2n}{n} S_1(2n) S_1(n) x^n =} \nonumber\\ &&
\frac{\displaystyle 1}{\displaystyle \sqrt{1-4 x}}
\Biggl\{
        4
        +4 x
        +\left[
                -4
                -\HA_0(x)
                +\text{G}\left(
                        \frac{\sqrt{1-4 \tau }}{\tau };x\right)
        \right] \sqrt{1-4 x}
        -3 \HA_{\frac{1}{4}}(x)
        -2 \HA_0(x)
\nonumber\\ && 
        +\frac{1}{2} \HA_0(x)^2
        +2 \HA_{\frac{1}{4},\frac{1}{4}}(x)
        +\frac{3}{2} \HA_{\frac{1}{4},0}(x)
        +\frac{3}{2} \HA_{0,\frac{1}{4}}(x)
        +2 \text{G}\left(
                \frac{\sqrt{1-4 \tau }}{\tau };x\right)
        -\text{G}\left(
                \frac{1}{\tau },\frac{\sqrt{1-4 \tau }}{\tau };x\right)
\nonumber\\ &&
        -6 \text{G}\left(
                \frac{1}{1-4 \tau },\frac{\sqrt{1-4 \tau }}{\tau \
};x\right)
        +\frac{1}{2} \text{G}\left(
                \frac{\sqrt{1-4 \tau }}{\tau },\frac{1}{\tau };x\right)
        -\frac{1}{2} \text{G}\left(
                \frac{\sqrt{1-4 \tau }}{\tau },\frac{\sqrt{1-4 \tau \
}}{\tau };x\right)
\Biggr\}.
\nonumber\\
\end{eqnarray}
The number of terms quickly grows for higher powers in $\mu$, but the above example 
shows that closed-form expressions using {\rm G}-functions at main argument $x$ can be
found. In general, the $\mu$ expansion does not substitute the complete result, however.
\section{\boldmath $\mu$-deformed $\zeta$-values from a modified $\mu$-polylogarithm} 
\label{sec:C}

\vspace*{1mm}
\noindent
Modified $\mu$-polylogarithms allow for a direct calculation of the associated 
$\mu$-deformed $\zeta$-values for $x \rightarrow 1$. We consider
\begin{eqnarray}
\label{MOD}
\tilde{\Li}_s(x;\mu) = \sum_{k=1}^\infty \frac{x^k}{k^s} \frac{1}{1+k \mu}
\end{eqnarray}
of Ref.~\cite{Gavrilik:2018cgj}. Here the $\mu$-dependent part introduces a Hurwitz-type
structure, cf.~Ref.~\cite{HURWITZ}. Resumming  (\ref{MOD}) for $x=1$ one obtains
\begin{eqnarray}
\tilde{\zeta}_1(\mu) &=& \psi\left(1+ \frac{1}{\mu}\right) + \gamma_E, \\
\tilde{\zeta}_2(\mu) &=& \zeta_2 - \mu \left[\psi\left(1+ \frac{1}{\mu}\right) + 
\gamma_E\right], 
\\
\tilde{\zeta}_3(\mu) &=& \zeta_3 - \mu \zeta_2 + \mu^2 \left[\psi\left(1+ 
\frac{1}{\mu}\right) 
+\gamma_E \right], \\
\tilde{\zeta}_4(\mu) &=& \zeta_4 - \mu \zeta_3 + \mu^2 \zeta_2 -
 \mu^3 \left[\psi\left(1+ \frac{1}{\mu}\right) + \gamma_E 
\right],
\end{eqnarray}
etc. Here $\gamma_E$ denotes the Euler-Mascheroni constant and $\psi(z)$ the 
Digamma-function. In this modified definition also $\tilde{\zeta}_1(\mu)$ exists, but 
it diverges logarithmically for $\mu \rightarrow 0$, while for $s > 1$ the usual 
$\zeta$-values are obtained.

\vspace*{5mm}
\noindent
{\bf Acknowledgment.}~~We thank Kolleg Mathematik Physik Berlin for financial support 
and P.~Marquard, C.~Schneider and K.~Sch\"onwald for discussions. The research of A.M.G. 
was also funded by the National Academy of Sciences of Ukraine by its priority project 
No.~0122U000888.


\end{document}